\documentclass[unnumsec,webpdf,modern,large]{oup-authoring-template}

\onecolumn 
\setcitestyle{numbers,square,comma}

\theoremstyle{thmstyleone}%
\theoremstyle{thmstyletwo}%
\theoremstyle{thmstylethree}%

\usepackage{algorithm}
\usepackage{algpseudocode}
\usepackage{longtable}
\usepackage{booktabs}
\usepackage{amsmath}
\usepackage{xcolor}
\usepackage{hyperref}
\usepackage{multirow} 
\newcommand{\Hsquare}{%
  \text{\fboxsep=-.2pt\fbox{\rule{0pt}{1.5ex}\rule{1.5ex}{0pt}}}%
}
\usepackage{makecell}

\usepackage[caption=false,position=bottom,labelformat=simple,labelsep=space]{subfig}
\newcommand{\red}[1]{\textcolor{red}{#1}}

\renewcommand{\textcolor}[2]{#2}

\usepackage{ulem}
\newcommand\crossout{\bgroup\markoverwith{\textcolor{cyan}{\rule[0.5ex]{1pt}{1.5pt}}}\ULon}
\renewcommand{\crossout}[1]{\unskip}

\begin{document}

\firstpage{1}


\title[Effectiveness of Privacy Transparency Channels]{Exploring the Effectiveness of Google Play Store's Privacy Transparency Channels}

\author[1]{Anhao Xiang\ORCID{0000-0002-8850-8583}}
\author[2]{Weiping Pei\ORCID{0000-0002-6831-2313}}
\author[1,$\ast$]{Chuan Yue\ORCID{0000-0002-6095-4768}}

\authormark{Xiang et al.}

\address[1]{\orgdiv{Department of Computer Science}, \orgname{Colorado School of Mines}, \orgaddress{\street{1500 Illinois St}, \postcode{80401}, \state{Colorado}, \country{United States}}}
\address[2]{\orgdiv{School of Cyber Studies}, \orgname{The University of Tulsa}, \orgaddress{\street{800 S. Tucker Drive Tulsa}, \postcode{74104}, \state{Oklahoma}, \country{United States}}}

\corresp[$\ast$]{Corresponding author. Colorado School of Mines, Department of Computer Science, 1500 Illinois St, 80401, Colorado, United
States. Email:~\href{email:chuanyue@mines.edu}{chuanyue@mines.edu}}

\abstract{With the requirements and emphases on privacy transparency placed by regulations such as GDPR and CCPA, the Google Play Store requires Android developers to more responsibly communicate their apps' privacy practices to potential users by providing the proper information via the \textit{data safety}, \textit{privacy policy}, and \textit{permission manifest} privacy transparency channels. However, it is unclear how effective those channels are in helping users make informed decisions in the app selection and installation process. In this article, we conducted a study for 190 participants to interact with our simulated privacy transparency channels of mobile apps. We quantitatively analyzed (supplemented by qualitative analysis) participants' responses to five sets of questions. We found that data safety provides the most intuitive user interfaces, privacy policy is most informative and effective, while permission manifest excels at raising participants’ concerns about an app's overall privacy risks. These channels complement each other and should all be improved.

}


\maketitle

\thispagestyle{empty}

\section{Introduction}
\label{sec:introduction}

Many mobile apps are inconsiderate, careless, aggressive, and even abusive in their practices on sensitive user data~\cite{zhoupolicycomp, zimmeck2019maps, bui2021consistency, feal2020angel,lehman2022hidden, reardon201950}.
Therefore, major app stores have been continuously improving their interfaces to 
require developers more responsibly communicate their apps' privacy and security practices to potential users during the \textit{app selection and installation process}.
Many of those improvements aimed specifically to increase privacy transparency and assist users in assessing privacy risks prior to making an app installation decision; they comply with the requirements and emphases placed by regulations such as the General Data Protection Regulation (GDPR)~\cite{GDPR_2024} and the California Consumer Privacy Act (CCPA)~\cite{CCPA_2024} 
on informing users about the details of the collection and sharing of their data and the rights they have.
GDPR is a European Union law enacted in 2016 that protects EU citizens' fundamental right to privacy, while CCPA is a California state statute enacted in 2018 that grants residents rights to access, delete, and opt out of the sale of their personal information.

\subsection{Three Privacy Transparency Channels, Their Roles, and Knowledge Gap}
\label{sec:Google Play Store and  Privacy Transparency Channels}

On the Google Play Store~\cite{Google_Play}, Android app users can learn about apps' privacy practices~(e.g., data collection, sharing, and storage) mainly through the \textit{\textbf{data safety}}, \textit{\textbf{privacy policy}}, and \textit{\textbf{permission manifest}} interfaces (as exemplified in Figure~\ref{fig:channels_examples} in Appendix~\ref{sec:appendix_screenshots}).
We refer to these three interfaces as three \textit{\textbf{privacy transparency channels}} in this study.

\textbf{Permission manifest channel.}
For an app, a technical-oriented permission manifest (file) has been required since the inception of the Android system.
Apps may require to access a range of system permissions to support their functionality, and those permissions can have different impacts on users' privacy. 
The Google Play Store requires app developers to declare all needed permissions along with the brief descriptions of the capabilities of these permissions in the ``App permissions'' page (as exemplified in Figure~\ref{fig:permission_example}). 
A user needs to click on the right arrow symbol in the ``About this app'' section of an app's main page and come to the app's detailed description page in order to further access the ``App permissions'' page (via a ``See More'' link).

\textbf{Privacy policy channel.}
As the legal binding between app developers and app users, a privacy policy (as exemplified in Figure~\ref{fig:privacy_policy_example}) has been required for an app since 2016. It typically contains more privacy practice information of an app 
such as data retention and transfer practices, legal basis for data processing, and users' data rights.
App developers are required to include at the end of the detailed data safety page (of the data safety channel) a URL linking to the app’s privacy policy webpage, which is typically hosted on the app developers' website and can be viewed by a user on a mobile browser.

\textbf{Data safety channel.}
A new and more user-oriented data safety section has been required on the Google Play Store since 2022, with its interface design inspired in part by the privacy nutrition label related studies such as~\cite{kelley2009nutrition}.
This new channel includes an interface for an app's \textit{data safety summary section} listed below the ``About this app'' section of the app's main page (as exemplified in Figure~\ref{fig:data_safety_summary_example}); it further includes an interface for the app's \textit{detailed data safety page} (as exemplified in Figure~\ref{fig:data_safety_example}) accessible via either the ``See details'' link or the right arrow symbol in the data safety summary section. 
The detailed data safety page allows developers to provide information related to data shared, data collected, and security practices.
More importantly, developers should follow Google's Android developer guide~\cite{goolePlayStoreDataSafetyDetails} to specify the detailed types of data collected or shared, along with the respective collection and sharing purposes.

These three channels offer information in different ways to enhance users' privacy risk awareness.
While privacy practice information may also appear in app reviews, Nema et al. found that privacy-related reviews constitute less than 1\% of 287 million reviews from two million apps across 29 categories~\cite{nema2022analyzing}.

\textbf{Knowledge gap.}
However, these three channels are far from being adequate to achieve their privacy transparency goal, as their utility is hindered by 
many specific issues identified in them (Section~\ref{sec:related_work}).
Moreover, there is only very limited research on the effectiveness of some individual privacy transparency channels (Section~\ref{sec:related_work}); especially, no study has compared these
channels in terms how effective they are in helping users
understand apps' privacy practices and assess associated
risks as well as how they impact users' privacy concerns. 
With the data safety channel designed as the most prominent means of conveying an app's privacy practices, the privacy policy and permission manifest channels became less accessible for users.
Meanwhile, many users may not explore all channels.
For example, many users do not read privacy policies~\cite{kelley2009nutrition}, or do not read a privacy policy in its entirety~\cite{gluck2016short}. 
Comparing these channels can help identify if they are complementary and should be leveraged collectively to achieve better privacy transparency. 

\vspace{-1pt}
\subsection{Online User Study, Research Questions, and Contributions}
\label{sec:user_study_RQs}

In this article, we design and conduct an online user study (n=190, recruited via Prolific~\cite{Prolific}) in the United States. Each participant plays the role of an app user to (1) interact with the three privacy transparency channels of an Android app
(following a randomized channel interaction sequence, with the app randomly assigned from four apps with different levels of privacy sensitivity and privacy practice disclosure extensiveness) 
simulated on our website, and (2) respond to five sets' of built-in questions (pre-study, pre-interaction, during-interaction, post-interaction, and post-study).
Specifically, we aim to solicit the responses from participants for us to quantitatively (supplemented by qualitative analysis) answer three research questions.
\textbf{RQ1:} What are users’ understanding of an app’s privacy practices as conveyed through the three privacy transparency channels for the app?
\textbf{RQ2:} How do the privacy practices conveyed through the three privacy transparency channels impact users’ judgment of the risks associated with installing and using an app?
\textbf{RQ3:} What are users' opinions on the three privacy transparency channels regarding the provided information, user interfaces, and overall effectiveness?
We compare participants’ responses 
from the app category, privacy practice disclosure extensiveness, and channel sequence aspects
(\textbf{\textit{three main between-subjects factors}}). We also compare participants’ responses before and after interacting with
the three channels (\textbf{\textit{the main within-subjects factor}}) to assess changes in their understandings and perceptions.
Answering these three RQs with the detailed between-subjects and within-subjects comparisons (based on quantitative statistical tests, supplemented by qualitative thematic analysis) can help us collectively identify the strengths and weaknesses of the three channels, and derive channel improvement suggestions for enhancing users' awareness and understanding of privacy risks.

Based on participants' responses, we found that 
(1) all three channels are effective in increasing participants' overall understanding of an app's privacy practices, (2) the data safety and permission manifest channels are particularly effective in raising participants' concerns about the overall privacy risks of an app, and (3)  each channel has its own pros and cons.
Overall, we make three major contributions in this article:
(1) we built a web application that allows participants to interact with the simulated privacy transparency channels of apps, enabling us and future researchers to flexibly conduct relevant user studies (released via a GitHub repository~\cite{github_link});
(2) we designed and conducted an interaction study to answer the three RQs;
(3) we analyzed participants' responses to derive findings and provide recommendations to stakeholders including app store providers, developers, and users.

The rest of the article is organized as follows.
Section~\ref{sec:related_work} introduces the related work.
Section~\ref{sec:design_of_study} describes the design of our study.
Section~\ref{sec:results_analysis} presents and analyzes the results of our study.
Section~\ref{sec:discussion} discusses the implications of our key findings, provides recommendations to stakeholders, and discusses the limitations of this work and the potential
future work.
Section~\ref{sec:conclusion} concludes the article.

\section{Related Work}
\label{sec:related_work}

\textbf{Ongoing issues in privacy transparency channels.}
Prior studies have identified many specific issues in individual privacy transparency channels. 
For instance, researchers in~\cite{chen2023investigating, tang2021defining, ibdah2021should} examined users' understanding of technical terms in privacy policies, revealing that many users struggle with those terms, often due to a general lack of explanations for the terminology and unfamiliar terms.
Chen et al.~\cite{chen2021fighting} explored users' comprehension of websites' privacy policies in the post-CCPA era, discovering that wording variations led to users' different interpretations of a company's data practices and their own privacy rights. 
Wagner et al.~\cite{wagner2023privacy} revealed that privacy policies often contain vague descriptions of data practices and fail to specify users' opt-in choices for implicitly collected data, such as online activities. 
Khandelwal et al.~\cite{khandelwal2023unpacking} revealed some problems of over-, under-, and inconsistent-reporting of security or privacy practices in the data safety channel through a measurement study on 1.1M Android apps; 
they also looked into app developers' process and challenges when working with the data safety channel. 
Xiao et al.~\cite{Xiao_USEC_2023} revealed the prevalence of the inconsistency between iOS apps' privacy labels and data flows.
Arkalakis et al.~\cite{arkalakis2024abandon} reported the inconsistencies between the privacy practices declared in Android apps' data safety sections and apps' actual behaviors.
Malki et al.~\cite{malki2024exploring} identified inconsistencies and conflicting information across 20 popular female  mobile health Android apps' privacy policies, data safety sections, and app interfaces.
Studies such as~\cite{zhoupolicycomp, zimmeck2019maps} revealed similar problems of overbroad or inconsistent reporting of data practices in apps' privacy policies.
Researchers in~\cite{cao2021large,shen2021can,tahaei2023stuck} investigated the Android and iOS runtime permission model (i.e., dynamic contextual permissions are requested when users perform specific in-app actions); they revealed issues especially the insufficient explanations provided to users for them to accurately estimate privacy risks and infer the scope of permission groups.
Tahaei et al.~\cite{tahaei2023stuck} further highlighted the challenges that app developers face, including the confusion over the scope of some permissions and third-party library requirements.
Researchers also explored tools that support developers. 
For example, Li et al. designed three annotation-based Android Studio IDE plugins Coconut~\cite{li2018coconut}, Honeysuckle~\cite{li2021honeysuckle}, and Matcha~\cite{li2024matcha} that can help developers more accurately and efficiently write privacy policies, implement in-app privacy notices (e.g., explanations for runtime permission requests), and generate data safety labels, respectively.

\textbf{Effectiveness of privacy transparency channels.}
Another line of research examines how an individual channel or tool affects users' understanding of privacy practices. 
For example, Gluck et al.~\cite{gluck2016short} investigated the effectiveness of privacy policies in helping users understand the privacy practices of fitness wearable devices; 
they found that privacy notices (especially in short form) effectively improve users' overall understanding of privacy practices compared to the cases when no policy is read.
Zhang et al.~\cite{zhang2022usable} investigated the usability of the iOS app privacy label design by interviewing 24 users;
they found that while iOS privacy labels have increased nearly half of users' understanding of apps' privacy practices, prominent issues such as a confusing structure, unfamiliar terminology, and a lack of integration with permission settings or controls remain. 
Zhang et al.~\cite{zhang2023privacy} further 
found that a significant percentage of users' actual privacy questions are not answered in today’s iOS and Android privacy labels. 
Recently, Lin et al.~\cite{lin2023data} compared Android and iOS privacy label designs in an interview study; they found that while having differences, both designs share some same or similar problems previously identified for iOS labels. 
In studies such as~\cite{farke2021privacy, herder2020privacy}, researchers examined the effectiveness of privacy transparency tools (instead of channels for apps) like Google's My Activity dashboard in influencing users’ concerns as well as perceptions of the risks and benefits associated with the data collection. 
\red{In other studies such as~\cite{kulyk2020has}, researchers examined the effectiveness of cookie disclaimers provided by websites in changing users’ attitudes and behavior toward cookie use.}

Overall, research on the effectiveness of the three privacy transparency channels provided by the Google Play Store is very limited, and no study has compared these channels. Our study aims to bridge this knowledge gap.

\section{Design of The Study}
\label{sec:design_of_study}

We answer our three RQs by designing and conducting an interaction study, in which participants play the role of app users to interact with Google Play Store's three privacy transparency channels simulated on our website for Android apps.
Taking the interaction study approach with the simulated channel interfaces provides a more dynamic and realistic experience for our participants, compared to taking a pure survey or interview study approach.
Meanwhile, compared to the traditional localized lab study approach, our interaction study can be flexibly conducted online, allowing us to easily recruit a large and diverse group of participants.

\begin{figure*}[t]
  \centering
  \includegraphics[width=0.8\textwidth]{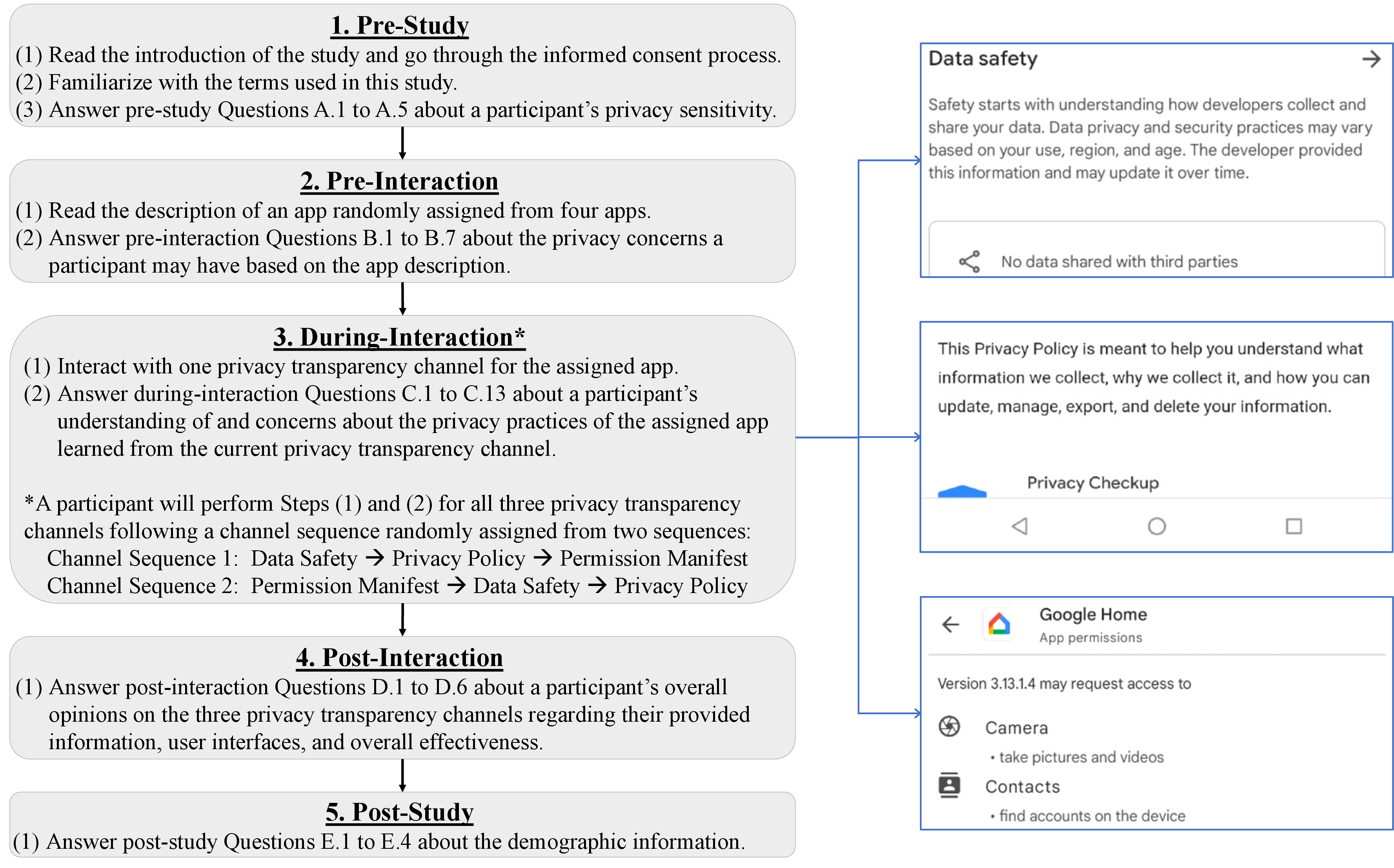}
  \caption{An Overview of the Interaction Study Procedure.}
  \label{fig:study_overview}
  \vspace{-10pt}
\end{figure*}

Figure~\ref{fig:study_overview} shows an overview of our interaction study procedure, which consists of five main stages conducted on our website. 
In the \textbf{\textit{Pre-Study}} stage, participants will read the introduction of the study and go through the informed consent process to decide whether they will participate in our study; they will read the terms (such as privacy transparency channels and privacy practices, \red{Section 7 of the Supplementary Material~\cite{github_link}}) used in this study; they will then answer Questions A.1 to A.5 (which are asked in this stage to avoid priming participants) regarding their privacy sensitivity characteristics.
In the \textbf{\textit{Pre-Interaction}} stage, participants will read the description of an app randomly assigned from four apps selected by us; they will then answer Questions B.1 to B.7 regarding the privacy concerns they may have based on the app description.
In the \textbf{\textit{During-Interaction}} stage, participants will interact with the three privacy transparency channels (\textbf{\textit{i.e., our main within-subjects factor}}) for the assigned app one at a time following a channel sequence randomly assigned from two sequences identified by us; 
immediately after interacting with each channel, participants will answer Questions C.1 to C.13 regarding their understanding of and concerns about the app's privacy practices.
Note that we only consider two channel sequences because on the Google Play Store the URL to the privacy policy appears at the end of the data safety page (Section~\ref{sec:Google Play Store and  Privacy Transparency Channels}).
In the \textbf{\textit{Post-Interaction}} stage, participants will answer Questions D.1 to D.6 about their overall opinions on the three channels regarding the provided
information, user interfaces, and overall effectiveness.
Finally in the \textbf{\textit{Post-Study}} stage, participants will answer Questions E.1 to E.4 about their demographic information.

\subsection{App Selection and Assignment}
\label{sec:Mobile App Selection and Assignment}
Prior study~\cite{nema2022analyzing} revealed that while users' privacy concerns on apps are not specific to any single app category on the Google Play Store, some categories may elicit privacy concerns from a larger number of users than others; for example, health-monitoring apps are generally more privacy-sensitive than news-feeding apps.
To minimize potential bias, we design our study with the app category (categories with low vs. high privacy sensitivity levels) as \textbf{\textit{the first main between-subjects factor}}. 
Furthermore, 
even the apps in the same category may disclose different amounts of
privacy practice information through the three channels, which may influence users' privacy concerns. 
Thus, we consider an app's privacy practice disclosure extensiveness (minimal vs. extensive) as \textbf{\textit{the second main between-subjects factor}}.
As a result, we set out to select four apps: two (one with minimal and the other with extensive privacy practice disclosure) from each of the two categories (one with low and the other with high privacy sensitivity) for our study.

\textbf{\textit{Category selection.}}
Nema et al.~\cite{nema2022analyzing} found that privacy-related reviews are predominantly negative, reflecting users' concerns about apps' privacy practices. 
Guided by the rankings of 29 categories (Table 3 in ~\cite{nema2022analyzing}) based on the ratio of privacy-related reviews to all reviews per category, we choose \textit{\textbf{News \& Magazines}} (ranked \#29) and \textit{\textbf{Dating}} (ranked \#1) as the categories with low and high privacy sensitivity levels, respectively, for our study.

\textbf{\textit{App selection.}}
To estimate apps' privacy practice disclosure extensiveness, 
we first obtain a list of the top-20 recommended apps under each category from the Google Play Store.
We then estimate the amount of privacy practice information disclosed on the three privacy transparency channels for each app. 
The term \textit{``privacy practices''} commonly refers to the collection and sharing of personal information by apps~\cite{andow2019policylint, bui2021consistency}; 
it also encompasses data retention and transfer, data rights, and user consent (e.g., consent withdrawal) practices, in the context of 
analyzing apps' privacy policies~\cite{xiang2023policychecker}.
Based on the estimation procedure detailed in Appendix~\ref{sec:appendix_Additional Design Details}, we choose two apps with minimal and extensive disclosures, respectively, from each category.
As a result, we selected four apps: Once, Bumble, DW, and USA TODAY, as shown in Table~\ref{tab:the_four_apps} for our study.

\begin{table*}[h]
\caption{The Selected Four Apps}
\scalebox{0.79}{
\begin{tabular}{|p{5.6cm}|p{6.3cm}|p{6.5cm}|}
\hline
\textbf{App Name}     & \textbf{Category (with Privacy Sensitivity Level)} & \textbf{Privacy Practice Disclosure Extensiveness} \\ \hline
\textit{\textbf{Once}: Find your Perfect Match}~\cite{googleOnceDating}      & Dating (High)       & Minimal        \\ \hline
\textit{\textbf{Bumble} Dating App: Meet \& Date}~\cite{googleBumbleDating}    & Dating (High)       & Extensive      \\ \hline
\textit{\textbf{DW}- Breaking World News}~\cite{googleDWNews}        & News \& Magazines (Low)          & Minimal        \\ \hline
\textit{\textbf{USA TODAY}: US \& Breaking News}~\cite{googleUSATODAYNews} & News \& Magazines (Low)         & Extensive       \\ \hline
\end{tabular}
}
\label{tab:the_four_apps}
\vspace{-5pt}
\end{table*}

\textbf{\textit{App assignment.}}
Each participant is randomly assigned with one of the four apps,
and will interact with the three channels following an interaction sequence (\textbf{\textit{i.e., the third main between-subjects factor}}) randomly assigned from the two that we identified.
Our website provides a participant with the app category information (in the \textit{Pre-Interaction} stage), but not the privacy sensitivity and practice disclosure extensiveness information (to avoid priming them).

\subsection{Simulated Privacy Transparency Channels}
\label{sec:Simulated User-Channel Interactions}
Our website hosts our PHP-based web application, which generates dynamic HTML webpages with JavaScript to simulate each app's main page, detailed description page, and user interfaces for the three channels. 
It integrates instruction prompts with the required steps to guide participants through structured interactions with those  interfaces. 
Participants need to click on the ``Next'' link in a prompt to reveal the next step and its instruction;
they need to click on the ``here'' link in the final prompt on a webpage to reveal and answer the questions designed for a corresponding stage.

In the \textit{Pre-Interaction} stage, our website displays two interfaces to a participant, replicating the main page and the detailed app description page of the assigned app on the Google Play Store.
Figure~\ref{fig:interface_app} in Appendix~\ref{sec:appendix_screenshots} exemplifies our simulated two interfaces of the USA TODAY app.

After familiarizing themselves with the assigned app,  
a participant will interact with the interfaces of the three 
channels one at a time in the \textit{During-Interaction} stage.
In particular, the interfaces for the data safety and permission manifest channels replicate the actual designs on the Google Play Store.
Meanwhile, the contents provided in these two interfaces are the same as those provided by the app developers on the Google Play Store. 
The privacy policy channel interface is a simple HTML webpage, similar to how the Google Play Store will let a user to view a privacy policy on a mobile browser. 
However, privacy policies are typically lengthy (e.g., Bumble's privacy policy contains over 8,000 words), and it would take a long time for a participant to carefully read a policy; therefore, we manually extract from each original policy the key sections and paragraphs that are most relevant to privacy practices, and present the condensed content in the privacy policy channel interface. Figures~\ref{fig:interface_data_safety},~\ref{fig:interface_privacy_policy}, and~\ref{fig:interface_permission_manifest} in Appendix~\ref{sec:appendix_screenshots} exemplify our simulated interfaces for the three channels of the USA TODAY app, respectively.
Their design details are provided in Appendix~\ref{sec:appendix_Additional Design Details}.

\subsection{The Five Sets of Questions}
\label{sec:Survey Questions}
We design five sets of questions (A.1 to A.5, B.1 to B.7, C.1 to C.13, D.1 to D.6, and E.1 to E.4) corresponding to the five stages of the study procedure (Figure~\ref{fig:study_overview}), and list them in Appendices~\ref{sec:appendix_Pre-Study Questions},~\ref{sec:appendix_Pre-Interaction Questions},~\ref{sec:appendix_During-Interaction Questions},~\ref{sec:appendix_Post-Interaction Questions}, and~\ref{sec:appendix_Demographic Questions}, respectively.
Question D.6 is open-ended and optional for general comments.
All other questions are close-ended (with most of them being Likert-scale questions) and required; meanwhile, seven of them (B.7, C.12, C.13, and D.1 to D.4) require a participant to provide open-ended explanations each with at least 15 words and with high quality.
Since we have three privacy transparency channels and we ask the identical (except for the channel name) C.1 to C.13 questions for each channel, a participant will answer 61 questions in total in our study.

\vspace{-5pt}
\subsubsection{Questions about participants' understanding of privacy practices}
\label{subsubsec:questions_for_RQ1}
We will mainly leverage during-interaction Questions C.2 to C.7 to answer our \textbf{\textit{RQ1}}. 
Question C.2 is about participants' overall opinions regarding whether based on the privacy practices they learned purely from a channel, their privacy-sensitive information could be collected, used, and potentially shared (with other parties) by the assigned app.
It is backed by the corresponding pre-interaction Question B.2, which is based on what participants learned from the app description.
Questions C.3 to C.7 are more detailed multiple-answer questions. Specifically, Questions C.3 and C.4 ask participants to select the types of privacy-sensitive information that the app may collect and use and may share with third parties, respectively;
Question C.5 asks participants to select the purposes for which the app may use their privacy-sensitive information;
Questions C.6 asks participants to identify the data rights they have regarding their privacy-sensitive information;
Question C.7 asks participants about their understanding of the app's data storage or retention practices.

\vspace{-5pt}
\subsubsection{Questions about participants' judgment of privacy risks}
\label{subsubsec:questions_for_RQ2}
We will mainly leverage during-interaction Questions C.8 to C.12, backed by the corresponding pre-interaction Questions B.3 to B.7, to answer our \textbf{\textit{RQ2}}.

\textbf{Questions on perceived risks.} 
Research in risk perception and communication~\cite{turner2011theory, skirpan2018s,gerber2019investigating} reveals that people often make decisions based on perceived rather than actual risks.
Therefore, we aim to assess how privacy transparency channels on the Google Play Store, as the means for communication, affect users' perceived online and physical risks associated with installing and using an app.
To this end, we include questions related to the perceived risks (Questions B.3 and B.4 and correspondingly C8 and C9) in both the \textit{Pre-Interaction} and \textit{During-Interaction} stages of our study.
We incorporate recommendations from the study on people's privacy risk perceptions~\cite{gerber2019investigating}
as well as the study on privacy risk categorization~\cite{karwatzki2017adverse}
into the design and phrasing of these four questions, reflecting some online and physical risks that may occur in real life.
Note that the only difference between the corresponding questions in Sets B and C is on the prefix phrases.
Questions in Set B begin with the prefix phrase: ``\textit{Based on what I learned so far from the app description, ...}'', 
while Questions in Set C begin with the prefix phrase: ``\textit{Based on the privacy practices I learned purely from the ... channel, ...}''.
Participants' perceived risks measured from the \textit{Pre-Interaction} stage 
will provide a baseline for us to analyze how their perceived risks measured from the \textit{During-Interaction} stage  
may change.

\textbf{Questions on perceived benefits.} 
Perceived risks alone may not dominate users' decision-making (e.g., on installing an app).
Thus, 
we design Questions B.5, B.6, C.10, and C.11 to measure if participants may perceive some benefits from an app's privacy practices.
Questions B.5 and C.10 are about the benefit of providing participants with a customized service tailored to their preferences or needs. Questions B.6 and C.11 are about the benefit of providing participants with enhanced online security protection such as better fraud detection or prevention. 

\textbf{Questions on the overall privacy risk concern  level.} 
We will further compare participants' responses to pre- and during- interaction Questions B.7 and C.12 to analyze how their overall privacy risk concerns on the assigned app would be influenced by the privacy practices conveyed through each channel.

\vspace{-5pt}
\subsubsection{Questions about participants' overall opinions on the three channels}
\label{subsubsec:questions_for_RQ3}
We will mainly leverage post-interaction Questions D.1 to D.5, backed by the during-interaction Question C.13, to answer our \textbf{\textit{RQ3}}.
Specifically, Question C.13 for each channel asks participants if the channel
 should be improved in some ways to help them better understand the privacy practices of the assigned app and better estimate the privacy risks associated with installing and using the app.
Questions D.1 to D.4 ask participants to rank the three channels in terms of the amount of information they provided, the clarity of the information they provided,
the intuitiveness of their user interfaces, and their overall effectiveness, respectively.
Question D.5 asks participants how in the future privacy transparency channels in general (instead of specific to a particular channel as in Question C.13) can be significantly improved in some ways.

\subsection{Participants and Ethical Considerations}
\label{sec:Participant Recruitment}

We used the Prolific~\cite{Prolific} crowdsourcing platform to recruit Android users as our participants and launch our study.
We only recruited adults residing in the United States due to the compensation requirement.

\textbf{Participants in the formal study.} 
We calculated the required sample sizes for the statistical tests to be performed in our study using the power analysis (as in \cite{mendel2017susceptibility, akter2020uncomfortable}).
To detect a medium effect size ($f$=$0.25$) with 80\% power in a one-way between-subjects test, 
the G*Power tool~\cite{faul2009statistical} suggests that we would need 45 participants in each group with 180 participants in total.
Therefore, we recruited 208 participants from Prolific in our formal study, but excluded 18 participants due to their very low response quality (i.e., they did not provide clear and relevant explanations
for their answers to \textbf{\textit{any of the seven questions}}: B.7, C.12, C.13, and D.1 to D.4).
Note that we only used participants' explanations to these seven questions for quality control.
We did not use attention checks or any other measures (e.g., forced breaks, completion time) for quality control.
The remaining \textit{\textbf{n=190}} participants are in four groups, with \textit{\textbf{50}}, \textit{\textbf{47}}, \textit{\textbf{47}}, and \textit{\textbf{46}} of them being assigned the Once, Bumble, DW, and USA TODAY apps, respectively (we refer to those participants as \textit{\textbf{Once~\#1-50}}, \textit{\textbf{Bumble~\#1-47}}, \textit{\textbf{DW~\#1-47}}, and \textit{\textbf{USA TODAY~\#1-46}}). 
Our formal study was conducted from March to early April 2024; the average and median time for completing the study was 61 and 53 minutes, respectively.
Each participant received \$8~USD as the compensation for completing the study (including the 18 excluded participants); 86.8\% of 190 participants further received the \$2~USD bonus due to their high-quality (in terms of clarity and relevance)  explanations for their answers to \textbf{\textit{all those seven questions}}.
The first two authors independently reviewed each participant's responses and held discussions to reach final decisions on excluding those 18 participants and issuing bonuses.
Note that the U.S. federal minimum wage is \$7.25/h. 
Based on the median completion time 39 minutes of our pilot study (Appendix~\ref{sec:appendix_Additional Design Details}), we increased the estimated completion time to 50 minutes for launching the formal study to consider potential varying time needs among a larger number of participants.
Prolific rated our pay at the ``Good'' level.
We did not receive any ``Underpaying'' warning from Prolific or any complaint from the 208 participants.

\textbf{Ethical considerations.} 
We received the IRB approval before recruiting any participant, and adhered to the approved IRB protocol to safeguard participants' privacy and security. 
We did not collect any \red{personally identifiable information} from them, and we securely stored their responses.
Prolific IDs were collected solely for the compensation purpose and were immediately deleted afterward. 

\section{Results and Analysis}
\label{sec:results_analysis}
In this section, we first present the demographic and privacy sensitivity information of our participants. 
We then analyze the overall and detailed results to answer our three RQs. 
Specifically, we analyze the detailed results based on the three main between-subjects factors and the main within-subjects factor.

We perform between-subjects comparison of ordinal data (e.g., responses to Likert-scale questions on perceived risks) by using the Wilcoxon rank sum test for two groups (e.g., news app vs. dating app participants); 
we perform between-subjects comparison of categorical data (e.g., the first place ranking of a channel in Set D questions) using the Chi-Square test for independence for two groups;
we perform within-subjects comparison (e.g., between a pair of corresponding Sets B and C questions) by using the Wilcoxon signed-rank test.
For each question involving more than 10 statistical tests, we control the False Discovery Rate (FDR) using the Benjamini–Hochberg (BH) procedure~\cite{benjamini1995controlling}, and report the $q$-values (adjusted from the original $p$-values) in APA format~\cite{APA}. 
The difference in response distribution is statistically significant if the $q$-value (or the $p$-value for a question without FDR control) is less than .05.

For questions that require open-ended explanations, the first two authors coded the responses by performing the thematic analysis~\cite{braun2006using}.
For each question, the two coders independently coded 20\% of responses before discussing and agreeing on an initial version of the codebook. Next, they independently coded all remaining responses and added new codes if necessary.
They then calculated the inter-coder reliability Krippendorff’s $\alpha$ coefficient~\cite{krippendorff2011computing,roth202112} for each question. 
The average and the standard deviation of all inter-coder reliability coefficients for all questions are 0.89 and 0.04, respectively, indicating a high inter-coder agreement level.
The two coders then discussed all codes to finalize the codebooks, 
which and all inter-coder reliability coefficients are provided in Section 1 of the Supplementary Material~\cite{github_link}.

\vspace{-5pt}
\subsection{Demographics and Privacy Sensitivity}
\label{subsec:Demographics_Privacy-Sensitivity}
\textbf{Demographic information.} Table~\ref{tab:demographic} in Appendix~\ref{sec:appendix_Additional Results} summarizes the demographics of our 190 participants. 
In brief, 76.8\% of them are between 18 and 49 years old; 
gender distribution is balanced between male (48.9\%) and female (47.4\%), with 3.7\% of participants self-identified as non-binary; 
52.6\% of participants hold Bachelor's, Associate, or equivalent degrees;  
78.9\% of participants are employed in over 20 different professions.

\textbf{Participants' privacy sensitivity characteristics.}
Participants were asked to select the types of information they considered privacy-sensitive based on their past experiences with apps (Question A.1 with its options adopted from the ``\textit{Data types and purposes}'' listed in Google's Android developer guide~\cite{goolePlayStoreDataSafetyDetails}). 
Financial information, personal information (e.g., name, email address), and photos and videos were the top-three types of sensitive information selected by 94.7\%, 91.1\%, and 87.9\% of participants, respectively, as shown in Figure~\ref{fig:A1} in Appendix~\ref{sec:appendix_Additional Results}.
Meanwhile, 91.1\% of participants \textit{\textbf{agreed (i.e., strongly or somewhat agreed)}} that some of their apps collect, use, and potentially share (with other parties) some types of privacy-sensitive information (Question A.2).
This result aligns with the responses from 85.3\% of participants who agreed that in general they are concerned about the overall privacy risks associated with apps (Question A.5).
These high levels of privacy awareness and concerns potentially explain why 89.5\% of participants agreed with the importance of knowing an app's privacy practices before deciding to install it (Question A.4).

Question A.3 asks participants whether they or someone they know had an experience of being a victim of some app related privacy violation; 74.2\% of participants reported having no such experience or were unsure.
Meanwhile, 52.6\%, 43.7\%, and 61.1\% of participants agreed that in the past they often reviewed the privacy practice information disclosed on the data safety, privacy policy, and permission manifest channels, respectively, before deciding to install an app (Question C.1). 
Only 10.0\%, 2.1\%, 8.5\%, and 10.9\% of the Once, Bumble, DW, and USA TODAY 
app participants, respectively, indicated that they installed and used the app before (Question B.1).

\subsection{RQ1: Understanding of Privacy Practices}
\label{sec:RQ1 users’ understanding of privacy practices}

We answer RQ1 by mainly analyzing during-interaction Questions C.2 (backed by the pre-interaction Question B.2) and C.3 to C.7 as we introduced in Section~\ref{subsubsec:questions_for_RQ1}.
\vspace{-5pt}
\subsubsection{Overall understanding}
\label{subsubsec:overall_B2_C2}
\begin{figure}[t]
  \centering
  \subfloat[Based on App Description (Question B.2).\label{fig:B2}]{
    \includegraphics[width=0.45\textwidth]{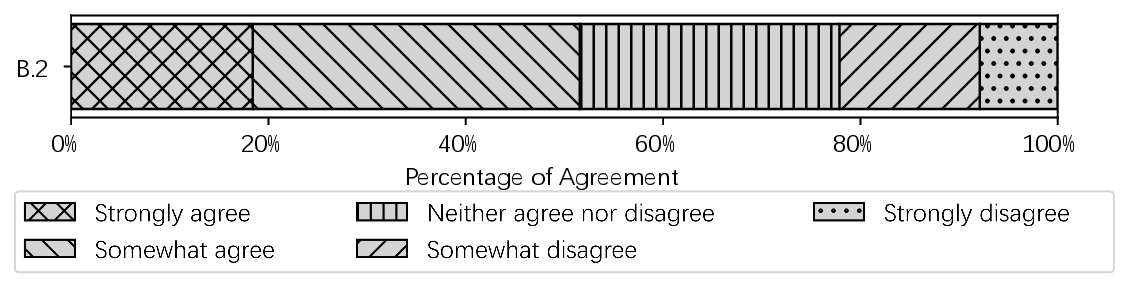}
  }\hfill
  \subfloat[Based on Interacting with Each Channel (Question C.2).\label{fig:C2}]{
    \includegraphics[width=0.49\textwidth]{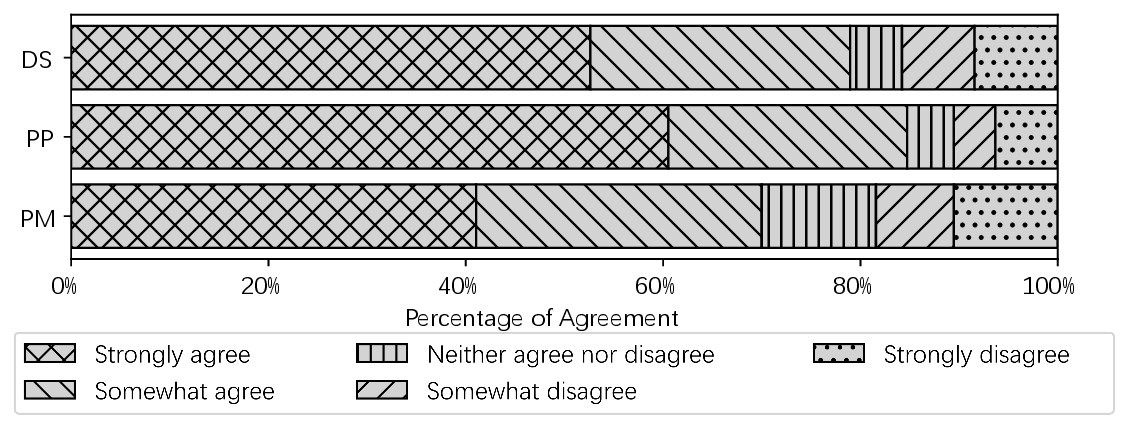}
  }
  \caption{Participants' Agreement Levels on An App's Privacy Practices Before (Question B.2) and After (Question C.2) Interacting with the Three Channels (n=190).}
  \label{fig:B2_C2}
  \vspace{-10pt}
\end{figure}

Overall, 51.6\% of the 190 participants agreed that their privacy-sensitive information could be collected, used, and potentially share by the assigned app based on the app description (Question B.2) as shown in Figure~\ref{fig:B2}.
More participants agreed with the corresponding statement in Question C.2 after they interacted with the data safety (\textbf{DS}), privacy policy (\textbf{PS}), and permission manifest (\textbf{PM}) channels, with the percentages increased to 78.9\%, 84.7\%, and 70.0\%, respectively, as shown in Figure~\ref{fig:C2}.
These increases are all statistically significant~($q<.001$), 
indicating that \textbf{\textit{privacy transparency channels are effective in increasing participants' overall understanding of an app's privacy practices}}.

\begin{figure*}[t]
  \centering
  \subfloat[App Category (B.2, C.2).\label{fig:C2-Category_Aspect}]{
    \includegraphics[scale=0.32]{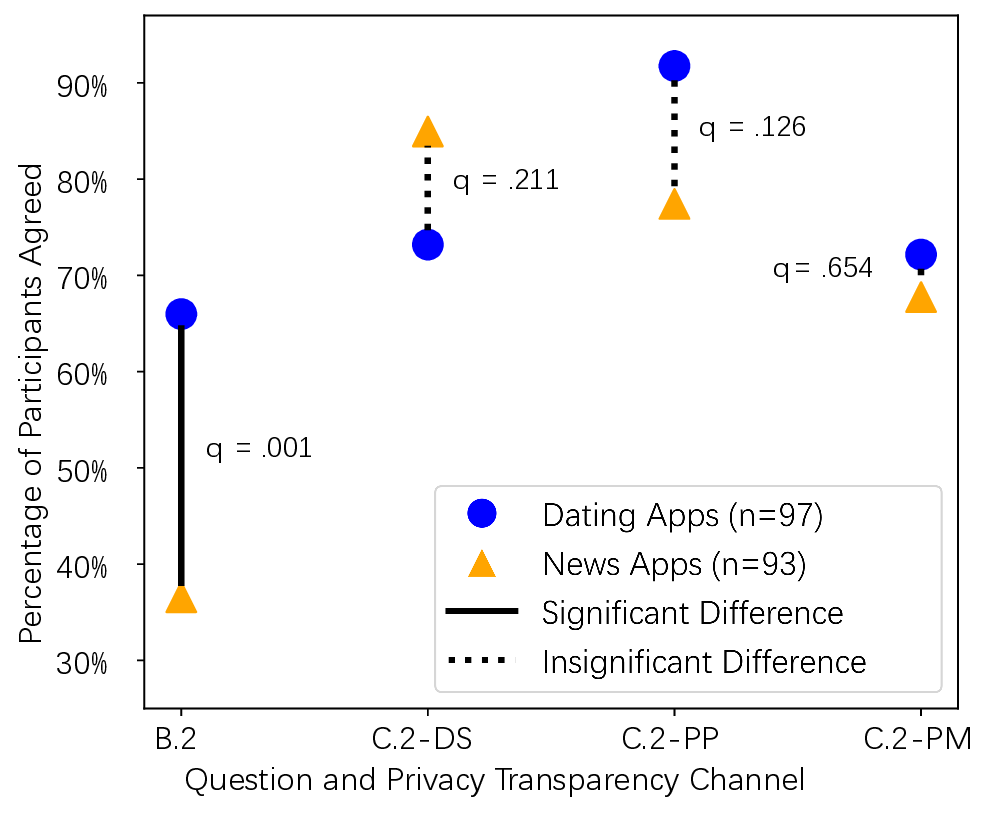}
  }\hfill
  \subfloat[App Disclosure Extensiveness (B.2, C.2).\label{fig:C2-Extensiveness_Aspect}]{
    \includegraphics[scale=0.32]{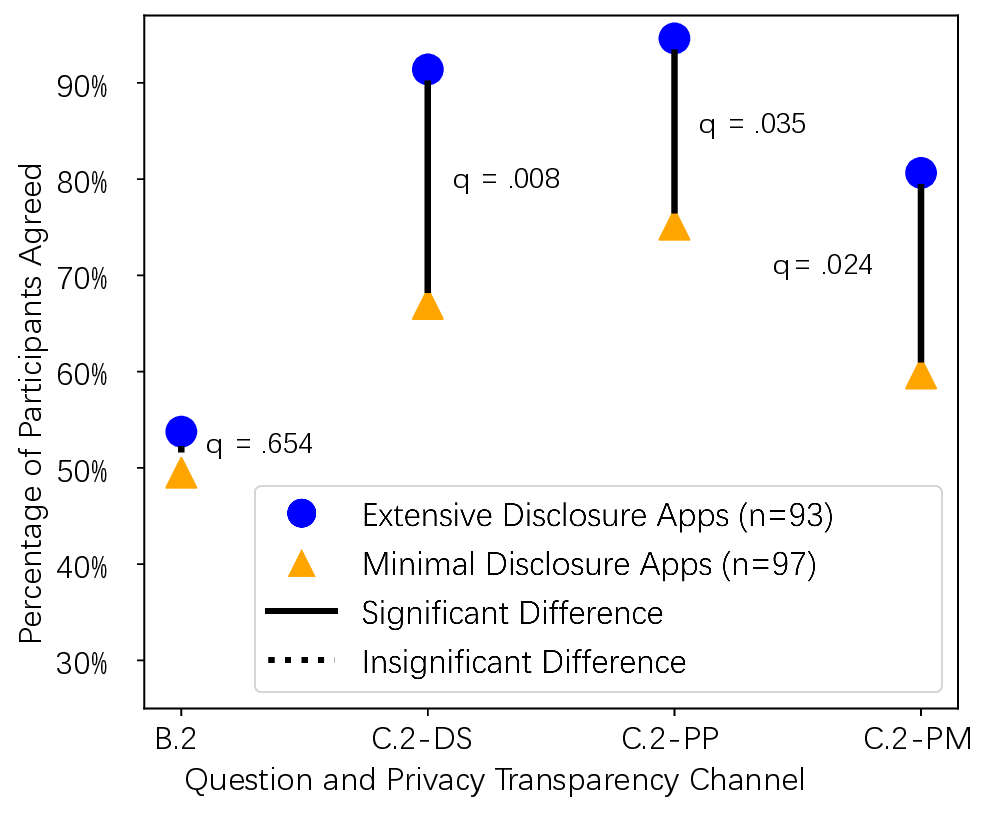}
  }\hfill
  \subfloat[Channel Sequence (B.2, C.2).\label{fig:C2-Sequence_Aspect}]{
    \includegraphics[scale=0.32]{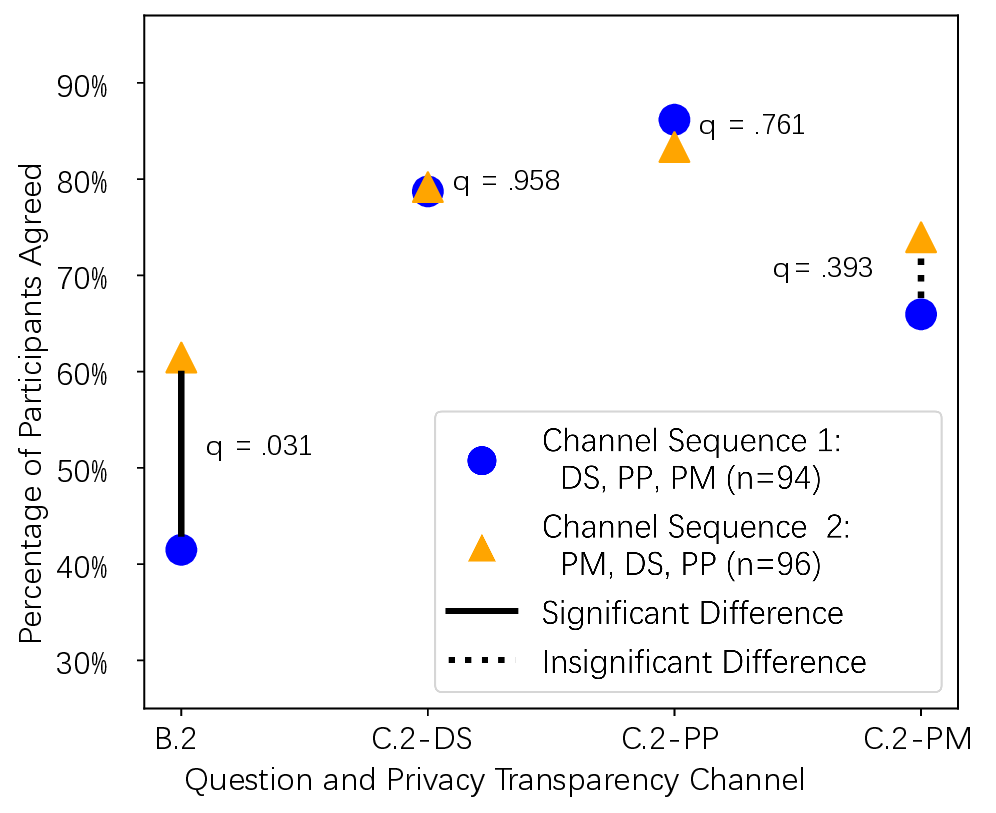}
  }
  \caption{Details of Participants' Agreement Levels in Questions B.2 and C.2 from Three Aspects.}
  \label{fig:C2_From_Three_Aspects}
\end{figure*}

We further analyzed the details from app category~(dating or news), disclosure extensiveness~(minimal or extensive), and channel interaction sequence~(\textbf{S-1}: DS$\rightarrow$PP$\rightarrow$PM; or \textbf{S-2}: PM$\rightarrow$DS$\rightarrow$PP) aspects.
The detailed results are shown in Figure~\ref{fig:C2_From_Three_Aspects}. 
Each subfigure corresponds to one aspect and shows the percentages of participants from the two groups who agreed with the statements in Questions B.2 and C.2. 
The agreement percentages for each question are depicted as two vertically aligned data points, 
which are used for a between-subjects comparison.
Subsequently, four between-subjects comparisons were conducted for each aspect. 
The results of these comparisons are illustrated in the figure with solid
and dashed lines indicating differences with or without statistical significance, respectively. 
Additionally, six within-subjects comparisons were conducted to
assess changes in participants' understandings and perceptions before and after interacting with different channels; 
these comparisons are not visualized in the figure to avoid cluttered lines.

From the app category aspect (Figure~\ref{fig:C2-Category_Aspect}), 65.9\% of the 97 dating app (Bumble or Once) participants whereas 36.6\% of the 93 news app (DW or USA TODAY) participants agreed with the statement in Question B.2; this difference is statistically significant ($q=.001$). 
Regarding the statement in Question C.2, dating app participants' interactions with the DS, PP, and PM channels led to the increases in agreement percentages to 73.2\%, 91.8\%, and 72.2\%, respectively; however, only the increase per the PP channel is statistically significant~($q<.001$).
For news app participants, corresponding percentages are increased to 84.9\%, 77.4\%, and 67.7\%, 
respectively, \textit{\textbf{all with statistical significance}}~($q<.001$); these significant increases 
are especially interesting because news apps are low on privacy sensitivity as we introduced in Section~\ref{sec:Mobile App Selection and Assignment}.
Moreover, the differences in agreement levels for Question C.2 between the dating and news app participants \textit{\textbf{are not statistically significant for all channels}}.

From the privacy practice disclosure extensiveness aspect (Figure~\ref{fig:C2-Extensiveness_Aspect}), 53.8\% of the 93 extensive disclosure app (Bumble or USA TODAY) participants while 49.5\% of the 97 minimal disclosure app (Once or DW) participants agreed with the statement in Question B.2; this difference is not statistically significant.
Extensive disclosure app participants' interactions with the DS, PP, and PM channels led to dramatic increases in Question C.2 agreement percentages to 91.4\%, 94.6\%, and 80.6\%, 
respectively, \textit{\textbf{all with statistical significance}}~($q<.001$);
while all three channels also increased the percentages of minimal disclosure app participants who agreed with the statement, 
only the 25.8\% and 17.5\% increases from the PP and DS channels are statistically significant~($q=.001$ and .035).
Moreover, the differences in agreement levels for Question C.2 between the extensive and minimal disclosure app participants \textit{\textbf{are statistically significant for all channels}}.

From the channel sequence aspect~(Figure~\ref{fig:C2-Sequence_Aspect}), 41.5\% of the 94 S-1 participants whereas 61.5\% of the 96 S-2 participants agreed with the statement in Question B.2; this difference is statistically significant ($q=.031$).
S-1 participants’ interactions with the DS, PP, PM channels led to the increases in Question C.2 agreement percentages to 78.7\%, 86.2\%, and 66.0\%, respectively, \textbf{\textit{all with statistical significance}}~($q<.001$). 
For S-2 participants, the corresponding percentages are increased to 79.2\%, 83.3\%, and 74.0\%; however, only the increases per the DS and PP channels are statistically significant~($q=.016$ and $ .003$).
Moreover, the differences in agreement levels for Question C.2 between the S-1 and S-2 participants \textit{\textbf{are not statistically significant for all channels}}.

\subsubsection{Understanding of Detailed Privacy Practices}
\label{subsubsec:detailed_C2_to_C7}
\begin{figure*}[t]
\centering

\subfloat[Ground Truth for Questions C.3 to C.7 on Number of Applicable Options for Each App's Privacy Transparency Channels.\label{fig:C3_C7_Ground_Truth}]{
  \includegraphics[width=0.32\textwidth]{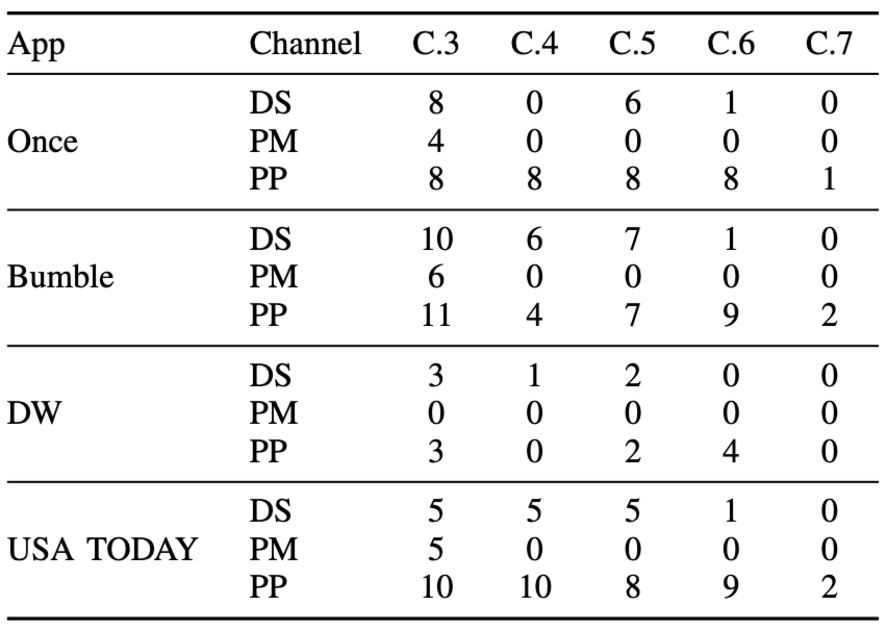}
}\hfill
\subfloat[Data Collection \& Use (Question C.3)\label{fig:C3}]{
  \includegraphics[width=0.32\textwidth]{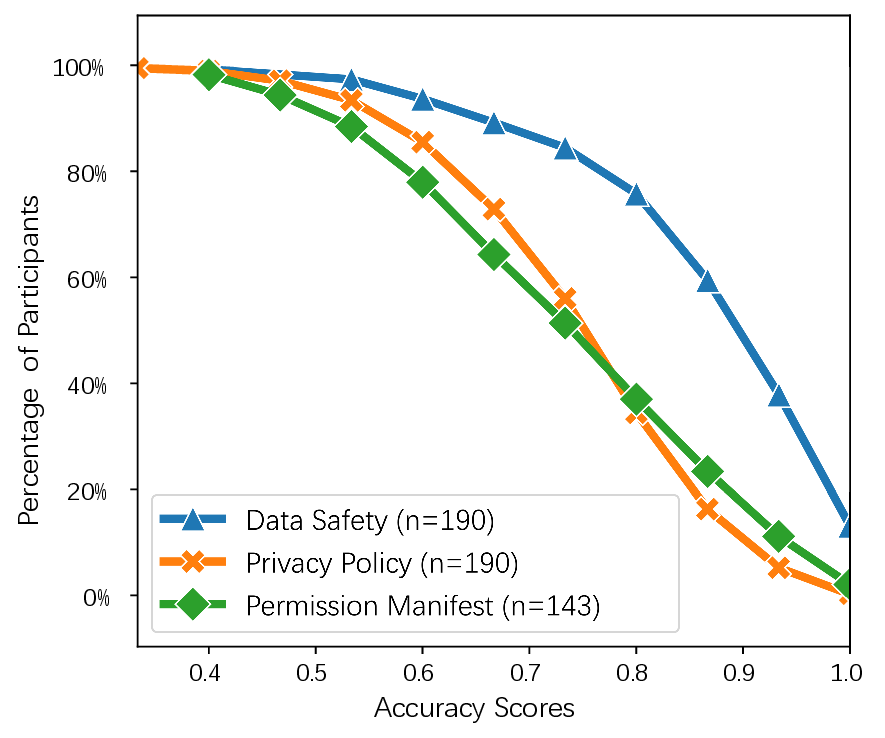}
}\hfill
\subfloat[Data Sharing (Question C.4)\label{fig:C4}]{
  \includegraphics[width=0.32\textwidth]{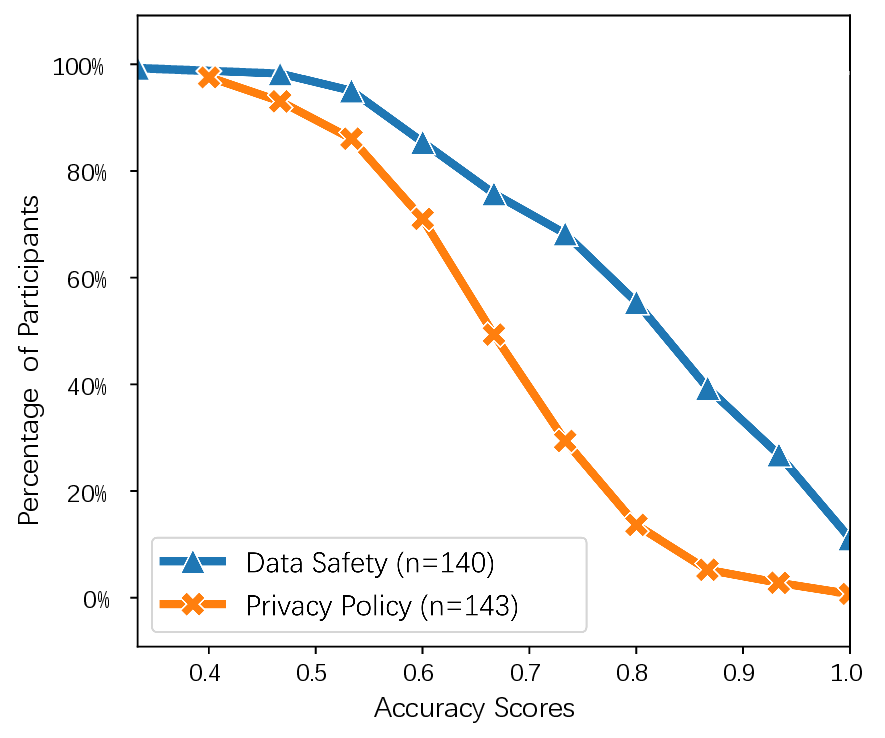}
}\\[8pt]

\subfloat[Data Use Purposes (Question C.5)\label{fig:C5}]{
  \includegraphics[width=0.32\textwidth]{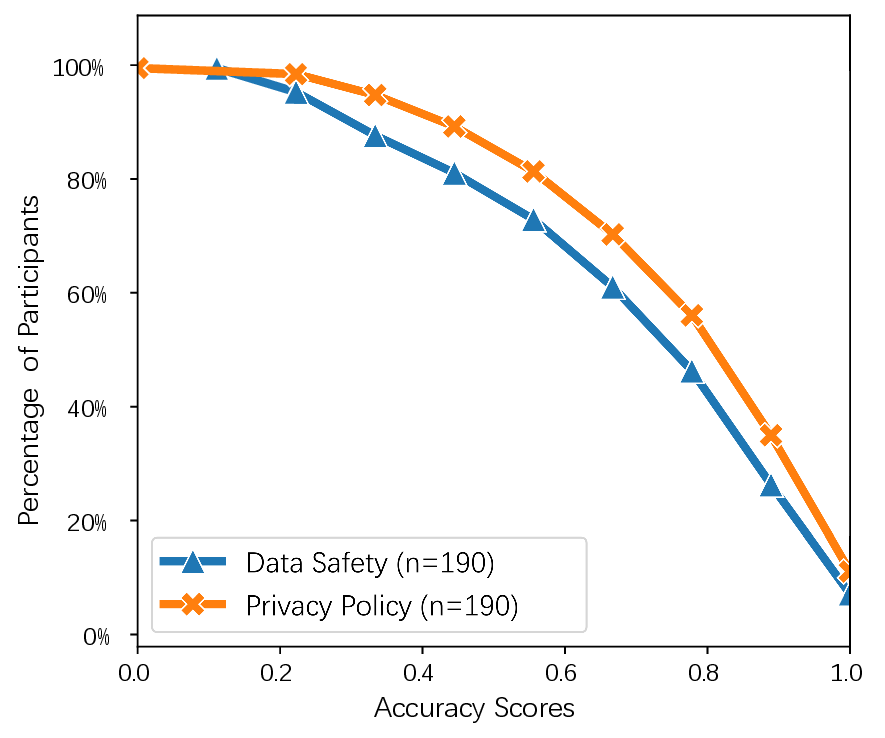}
}\hfill
\subfloat[Data Rights (Question C.6)\label{fig:C6}]{
  \includegraphics[width=0.32\textwidth]{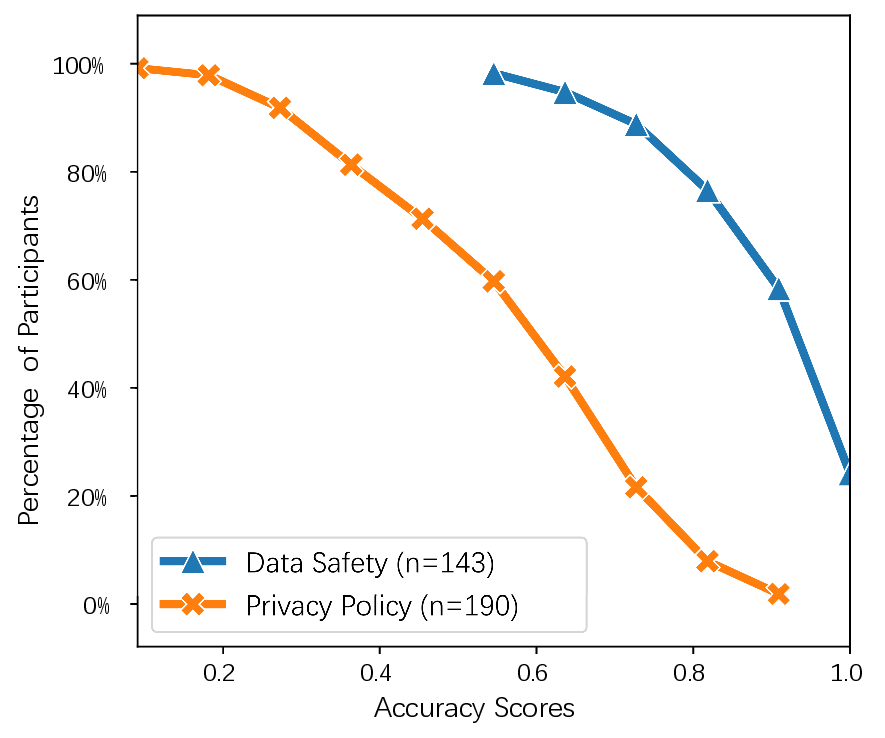}
}\hfill
\subfloat[Data Storage/Retention (Question C.7)\label{fig:C7}]{
  \includegraphics[width=0.32\textwidth]{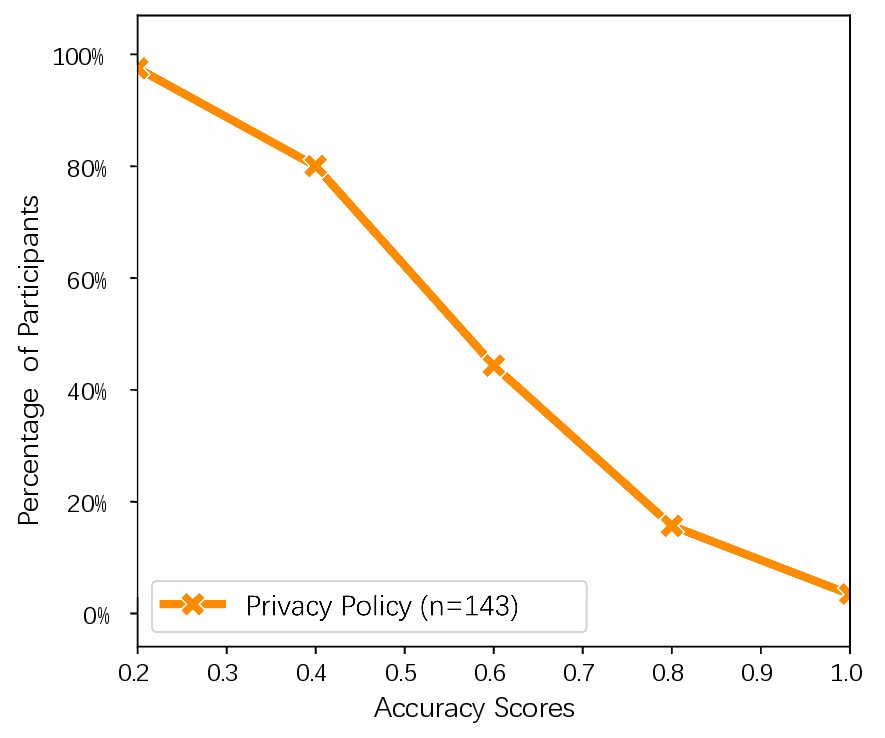}
}

\caption{Complementary Cumulative Distributions of the Response Accuracy for Questions C.3 to C.7. (A channel does not appear in a subfigure if it does not convey the corresponding practices in any of the four apps, n=143 is due to the exclusion of the DW app, and n=140 is due to the exclusion of the Once app, per the ground truth in
the subfigure 4a.)}
\label{fig:RQ1_CCDF}
\end{figure*}

Multiple-answer Questions C.3 to C.7 are about participants' understanding of the detailed privacy practices conveyed through each of the three channels.
The options for Questions C.3 and C.4 are the same as those for Question A.1 (Section~\ref{subsec:Demographics_Privacy-Sensitivity}).
The options for Questions C.5 to C.7 are constructed based on Google's Android developer guide~\cite{goolePlayStoreDataSafetyDetails} as well as GDPR~\cite{GDPR_2024} and CCPA~\cite{CCPA_2024} regulations.
To measure the extent to which participants captured the data practices conveyed through each channel of an app, we calculate the \textit{\textbf{accuracy}} of their responses; that is, we divide 
the sum of the total number of correctly selected options (i.e., TPs) and the total number of correctly unselected options (i.e., TNs) by the total number of options provided in a question.
To do so, we established the ground truth for each question regarding the applicable options 
for each app's privacy transparency channels as shown in Figure~\ref{fig:C3_C7_Ground_Truth}; zero means that a channel does not provide the information listed in a question (e.g., no data sharing information is in a permission manifest).

\textbf{Data collection and use, data sharing, and data use purposes.}
Figure~\ref{fig:C3} illustrates the complementary cumulative distributions of the response accuracy for Question C.3;
the percentages of participants who achieved 0.8 or greater accuracy in identifying data collection and use practices from the DS, PP, and PM channels are 75.8\%, 34.5\%, and 37.1\%, respectively.
It is interesting that DS and PM channels share a similar style of information presentation (i.e., using short labels and bullet lists), but the latter seems to be ineffective in helping participants establish the connection between an app's needed permissions and its data collection practices.
Part of this difference could be due to these two channels' different numbers of applicable options as shown in Figure~\ref{fig:C3_C7_Ground_Truth}.
As for Question C.4 (Figure~\ref{fig:C4}), only 13.6\% of participants achieved 0.8 or greater accuracy in identifying data sharing practices from the PP channel,
while the percentage from the DS channel is 55.4\%. 
Regarding Question C.5 (Figure~\ref{fig:C5}), 26.3\% of participants achieved 0.88 or greater accuracy in identifying data use purposes from the DS channel, while the percentage from the PP channel is 35.0\%.

\textbf{Data rights, and data storage or retention.}
Regarding Question C.6, Figure~\ref{fig:C6} shows that 76.6\% of participants achieved 0.8 or greater accuracy in identifying the only data right statement (which is on data erasure) provided in the DS channel of an app (excluding DW).
Although the privacy policies of all four apps provided rich information about data rights, only 7.9\% of participants achieved 0.8 or greater accuracy in identifying those rights.  
Only the PP channel provides information about data storage or retention practices (Question C.7); 
however, Figure~\ref{fig:C7} shows that only 15.7\% of participants were able to identify such practices from this channel with an accuracy of 0.8 or greater.

We also analyzed Questions C.3 to C.7 per the three between-subjects factors. 
We summarized the results in Appendix~\ref{sec:appendix_Additional Results}, and further provided the details in Section 2 of the Supplementary Material~\cite{github_link}.

\textbf{Key RQ1 takeaways or findings.}
First, all three channels 
are effective in increasing participants' overall understanding of an app's privacy practices, with the PP channel excelling at this; meanwhile, all three channels are effective especially to news app, extensive disclosure app, and S-1 participants.
Second, extensive app participants' agreement levels for Question C.2 are higher than that of minimal disclosure app participants for all channels.
Third, participants can more accurately capture the data collection and use practices from the DS channel than from the other two channels; the PM channel is limited to mainly convey such practices of an app.
Fourth, the DS channel is more effective than the PP channel in helping participants capture the data sharing practices of an app, but is less effective than the latter from the aspect of the data use purposes.
Fifth, although only the PP channel provides rich information for both data rights and data storage or retention practices, only a small percentage of participants accurately captured such information.

\subsection{RQ2: Judgment of Privacy Risks}
\label{sec:RQ2 users’ judgment of the privacy risks}

We answer RQ2 by mainly analyzing during-interaction Questions C.8 to C.12 (backed by pre-interaction Questions B.3 to B.7) as we introduced in Section~\ref{subsubsec:questions_for_RQ2}.
Table~\ref{tab:RQ2_participants_perceptions} in Appendix~\ref{sec:appendix_Additional Results} provides the details of 190 participants' responses to these questions.

\vspace{-5pt}
\subsubsection{Perceived risks, benefits, and concerns based on an app's description}
\label{subsubsec:B3_to_B7_results}
We combined participants' responses to Questions B.3 and B.4 to analyze the perceived risks, and Questions B.5 and B.6 to analyze the perceived benefits. Similarly, we combined the corresponding during-interaction questions (C.8 and C.9; C.10 and C.11).
Each question pair has an ``Excellent'' reliability level ($\alpha\geq0.9$) indicated by Cronbach's Alpha~\cite{cronbach1951coefficient}.
We found that based on an app’s description, 23.7\% and 56.6\% of participants agreed that the privacy practices of the assigned app may harm them by incurring some risks and may benefit them, respectively.

From Question B.7, we found that 71 (37.4\%) participants agreed with the statement that 
based on what they learned so far from the app description, they are concerned with the overall privacy risks associated with installing and using the assigned app.
This question requires participants to provide open-ended explanations to their answers.
Among those 71 participants, 
the top-two explanations are: 
21 mentioned that the lack of privacy-related information raised their concerns, 
e.g., \textit{``I am concerned about the privacy risks primarily because I dont have enough information to make an informed decision''} (Once~\#11), and
21 anticipated that their sensitive information would be collected or shared based on the app's functionality,
e.g., \textit{``Because I would have to provide information for the dating app in order to be matched, I would worry that there is a risk of a breach of personal information''} (Once~\#9);
among 84 participants who 
\textit{\textbf{disagreed (i.e., strongly or somewhat disagreed)}} 
with the statement, 
20 anticipated low privacy risks, 
e.g., \textit{``I think this app is not a high risk app that will require a lot of sensitive information so I am generally not concerned in that regard''} (USA TODAY~\#5), and
17 believed that no or limited sensitive information would be collected or shared,
e.g., \textit{``I am not concerned in this case because it is not collecting sensitive information such as name or financial information. The data is encrypted in transit. I don't care about location, app activities and performances.  That does not identify me.''} (DW~\#1).

\vspace{-5pt}
\subsubsection{Impact of channel interaction on perceived risks, benefits, and concerns}
\label{subsec:impact of channels on RQ2}
Questions C.8 to C.12 correspond to Questions B.3 to B.7, respectively, but are about what participants perceived after 
interacting with each channel.

\textbf{Impact on perceived risks.}
We found that 28.4\%, 32.8\%, and 33.1\% of 190 participants agreed that they may get harm from some risks based on the privacy practices they learned purely from the DS, PP, and PM channels, respectively.
Compared to the percentage of participants who perceived some risks based on the app description, all three channels increased the percentages, with 4.4\%, 9.1\%, and 9.4\%, respectively. 
The increases from the PP and PM channels are statistically significant~($q=$ .002 and .001).

\textbf{Impact on perceived benefits.}
We found that 51.3\%, 61.0\%, and 46.1\% of 190 participants agreed that the assigned app may benefit them based on the privacy practices they learned purely from the DS, PP, and PM channels, respectively.
Compared to the percentage of participants who perceived some benefits based on the app description, the DS and PM channels decreased the percentages, with -5.3\% and -10.5\%, respectively;
however, the PP channel increased the percentage by 4.4\%.
Only the decrease from the PM channel is statistically significant~($q=.007$).

We also analyzed the perceived risks and benefits per the three between-subjects factors.
We summarized the results in Appendix~\ref{sec:appendix_Additional Results}, and further provided the details in Section 3 of the Supplementary Material~\cite{github_link}.

\textbf{Impact on concerns about overall privacy risks.}
From Question C.12, we found that 94 (49.5\%), 89 (46.8\%), and 120 (63.2\%) participants agreed with the statements that based on 
the privacy practices they learned purely from the DS, PP, and PM channels, respectively, they are concerned with the overall privacy risks
associated with installing and using the assigned app.
Compared to the results from Question B.7, all three channels increased the percentages of participants who are concerned with the overall privacy risks, with 12.1\%, 9.6\%, and 25.8\%, respectively.
The increases from the DS and PM channels are statistically significant~($q = .008$ and .001), indicating that \textbf{\textit{these two channels are effective in raising participants’ concerns about the overall privacy risks of an app}};
the PM channel, despite being the oldest yet least emphasized channel (in recent years), excels at this.

Question C.12 also requires participants to provide open-ended explanations to their answers.
For \textbf{\textit{the DS channel, among 94 participants who agreed with the statement}}, the top-two explanations are: 
41 anticipated that their sensitive information would be collected or shared,
e.g., \textit{``I feel that installing the app would allow too much of my information to be shared with outside companies''} (USA TODAY~\#7), and
26 mentioned the lack of privacy transparency information such as explanations,
e.g., \textit{``They say they can share my device identifiers, but they don't say with who or why. That is a little concerning''} (DW~\#27);
\textbf{\textit{among 75 participants who disagreed with the statement}}, 
the top-two explanations are: 
18 believed that no or limited sensitive information would be collected or shared,
e.g., \textit{``The privacy risks with this app are minimal since data is not shared with a third party''} (Once~\#6), and
15 believed that the app's privacy practices are justified,
e.g., \textit{``I am not concerned about the overall privacy risks of using the DW app because it is clear to me that the app only collects information for analytics in order to improve the app itself''} (DW~\#2).
For \textbf{\textit{the PP channel, among 89 participants who agreed with the statement}}, the top-two explanations, from 30 and 17 participants, respectively, are 
the same as those for 
the DS channel;
\textbf{\textit{among 75 participants who disagreed with the statement,}} 26 mentioned sufficient transparency in privacy practices, 
e.g., \textit{``The detailed information provided about the data collected and the right of the user to access and have it deleted and removed, as well as not being shared with third party organizations allows the user to use the app with little or no concern due to the extreme transparency of DW''} (DW~\#35),
and nine believed that the app's privacy practices are justified,
e.g., \textit{``It seems that everything they do with my info benefits me one way or another. All the information they use, they do so to help with performance and utilization of the app''} (Bumble~\#44).
\textbf{\textit{For the PM channel, among 120 participants who agreed with the statement}}, the top-two explanations are: 40 mentioned excessive permissions requested beyond an app's functionality, 
e.g., \textit{``I am really concerned that it can access personal files on my device, photos, etc. Thats too much personal information for the benefits it gives''} (Once~\#43),
and 24 mentioned the lack of privacy transparency information such as explanations,
e.g., \textit{``The app asks for a lot of permissions without reason for these permissions. It does not detail how the the information is being used or why its being collected''} (Once~\#41); 
\textbf{\textit{among 53 participants who disagreed with the statement}}, 
10 believed
that the app’s permission requests are justified, 
e.g., \textit{``I am not too concerned as I feel like the information will help personalize the app for me specifically''} 
(USA TODAY~\#43),
and seven mentioned that they can manage permissions,
e.g., \textit{``I am not concerned with the privacy risks because I have the option to change or remove it whenever I like''} (Bumble~\#39).

\begin{figure*}[t]
\centering

\subfloat[App Category (B.7, C.12)\label{fig:C12-Category_Aspect}]{
  \includegraphics[width=0.32\textwidth]{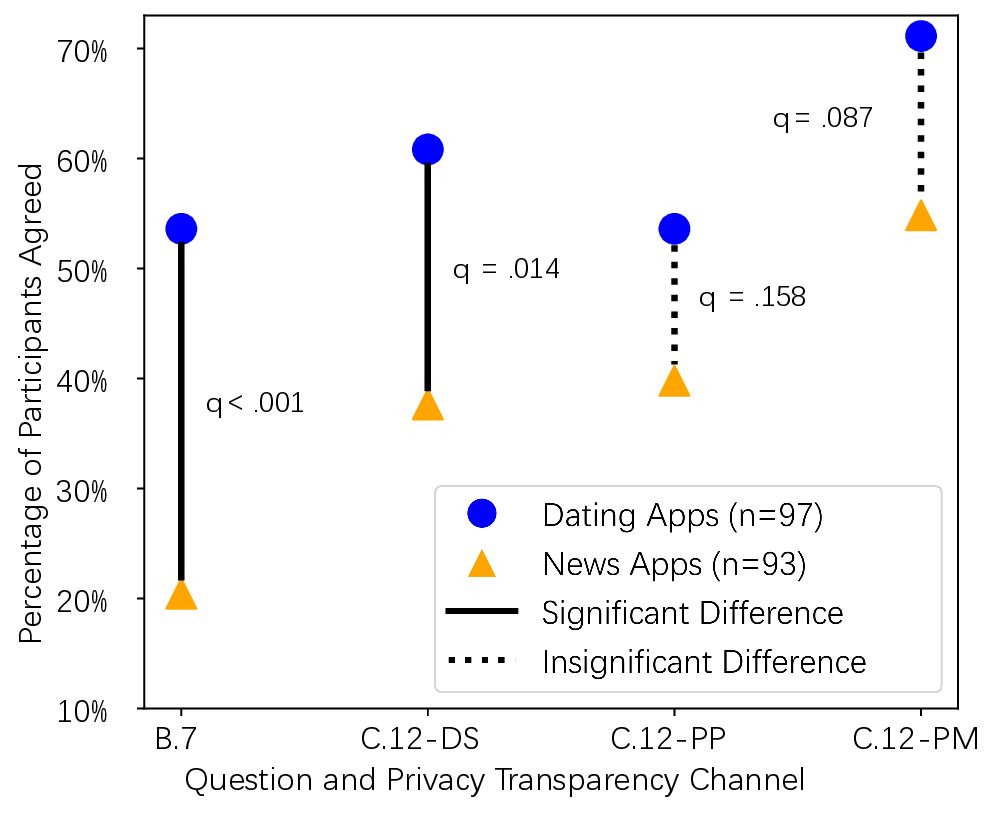}
}\hfill
\subfloat[App Disclosure Extensiveness (B.7, C.12)\label{fig:C12-Extensiveness_Aspect}]{
  \includegraphics[width=0.32\textwidth]{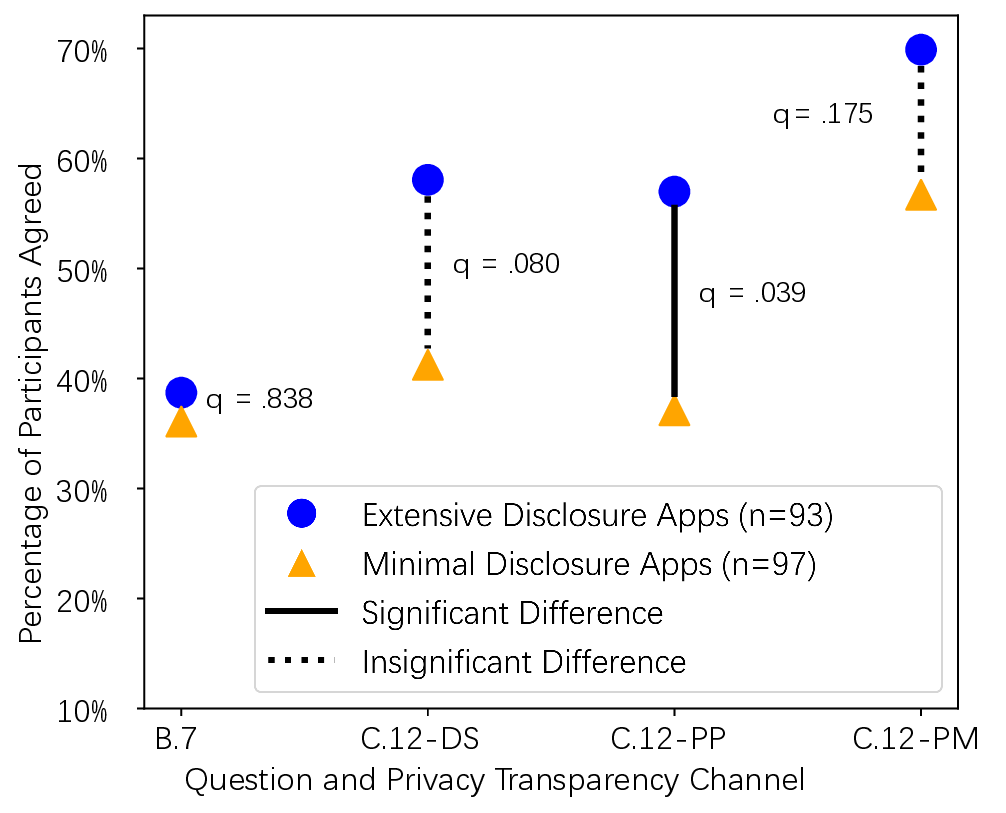}
}\hfill
\subfloat[Channel Sequence (B.7, C.12)\label{fig:C12-Channel_Sequence_Aspect}]{
  \includegraphics[width=0.32\textwidth]{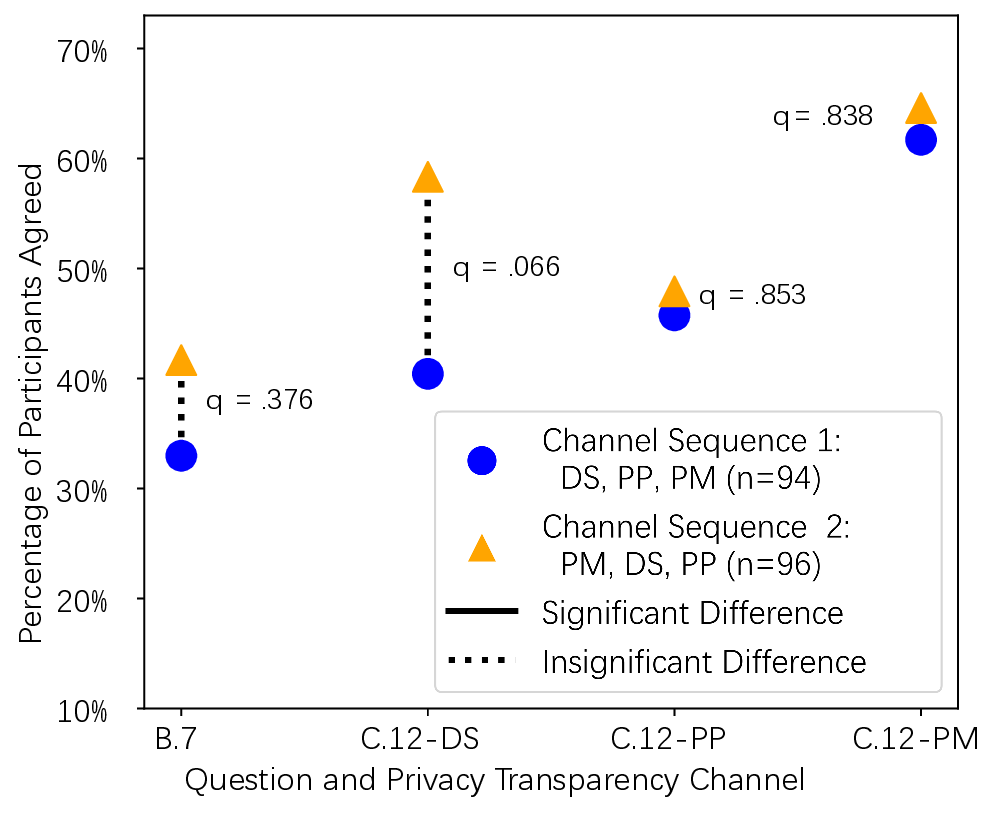}
}

\caption{Details of Participants' Agreement Levels in Questions B.7 and C.12 from Three Aspects.}
\label{fig:C12_From_Three_Aspects}
\end{figure*}

From the app category aspect~(Figure~\ref{fig:C12-Category_Aspect}), 53.6\% of the 97 dating app whereas 20.4\% of the 93 news app participants agreed with the statement in Question B.7; this difference is statistically significant ($q<.001$). 
Regarding the statement in Question C.12, dating app participants' interactions with the DS and PM channels led to the increases in agreement percentages to 60.8\% and 71.1\%, respectively; but only the latter increase is statistically significant~($q=.010$); interactions with the PP channel did not change the agreement percentage. 
For news app participants, corresponding percentages are increased to 37.6\%, 39.8\%, and 54.8\%, respectively, \textit{\textbf{all with statistical significance}}~($q=.010$, $.011$, and $< .001$).
Moreover, for the DS channel, the difference in agreement levels for Question C.12 between dating and news app participants is statistically significant.

From the privacy practice disclosure extensiveness aspect~(Figure~\ref{fig:C12-Extensiveness_Aspect}), 38.7\% of the 93 extensive disclosure app and 36.1\% of the 97 minimal disclosure app participants agreed with the statement in Question B.7; this difference is not statistically significant.
Extensive disclosure app participants' interactions with the DS, PP, and PM channels led to increases in Question C.12 agreement percentages to 58.1\%, 57.0\%, and 69.9\%, 
respectively, \textit{\textbf{all with statistical significance}}~($q=.002$, $.014$, and $<.001$). 
While all three channels also increased the percentages of minimal disclosure app participants who agreed with the statement, only the increase (20.6\%) from the PM channel is statistically significant~($q=.002$).
Moreover, for the PP channel, the difference in agreement levels for Question C.12 between extensive and minimal disclosure app participants is statistically significant.

From the channel sequence aspect~(Figure~\ref{fig:C12-Channel_Sequence_Aspect}), 33.0\% of the 94 S-1 and 41.7\% of the 96 S-2 participants agreed with the statement in Question B.7; this difference is not statistically significant.
The S-1 participants' interactions with the DS, PP, and PM channels led to the increases in Question C.12 agreement percentages to 40.4\%, 45.7\%, and 61.7\%, respectively; however, only the increase per the PM channel is statistically significant~($q<.001$).
For S-2 participants, the corresponding percentages are increased to
58.3\%, 47.9\%, and 64.6\%; the increases from the DS and PM channels are statistically significant~($q=.014$ and < .001).
Moreover, the differences in agreement levels for Question C.12 between S-1 and S-2 participants \textit{\textbf{are not statistically significant for all channels}}.

\textbf{Association analysis.}
We explored the potential associations among three groups of participants who agreed, disagreed, or were neutral with the statements in Questions B.7 and C.12.
We also explored the potential participant associations of the following variables with Questions B.7 and C.12, respectively: (1) participants’ overall understanding of an app’s privacy practices (Question C.2), and (2) their privacy sensitivity characteristics (Questions A.1 to A.5). 
The Chi-square tests found statistically significant association relationships between participants on Questions B.7 and C.12 as well as between participants on Questions C.2 and C.12.
The detailed test results are provided in Section~5 of the Supplementary Material~\cite{github_link}.
We avoided the association analysis for Question B.1 due to extremely skewed and small group sizes (Section~\ref{subsec:Demographics_Privacy-Sensitivity}).

\textbf{Key RQ2 takeaways or findings.}
First, the DS and PM channels are effective in raising participants' concerns about the overall privacy risks of an app, with the PM channel excelling at this. 
Participants' explanations to their answers to Question C.12 especially indicated that sensitive information collection or sharing and excessive permission requests are major reasons for their concerns.
Meanwhile, all three channels are effective especially to news app participants and 
extensive disclosure app participants.
Second, dating app participants' agreement levels for Question C.12 are higher than that of news app participants for the DS channel;
extensive app participants' agreement levels are higher than that of minimal disclosure app participants for the PP channel.
Third, 
the PP and PM channels clearly increased the percentages of participants who perceived some risks.
Fourth, the PM channel clearly decreased the percentages of participants who perceived some benefits.

\subsection{RQ3: Overall Opinions on Three Channels}
\label{subsec:RQ3_overall_opinions}
We answer RQ3 by mainly analyzing post-interaction Questions D.1 to D.5 (backed by the during-interaction Question C.13) as we introduced in Section~\ref{subsubsec:questions_for_RQ3}.

\begin{figure*}[t]
\centering

\subfloat[Rankings in Order of the \textit{Amount of Information}.\label{fig:D1}]{
  \includegraphics[width=0.47\textwidth]{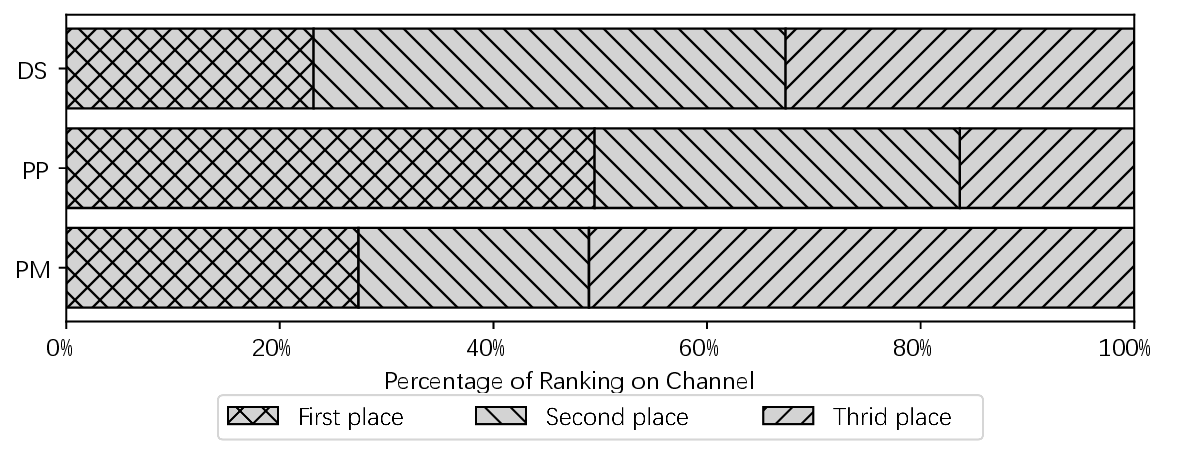}
}\hfill
\subfloat[Rankings in Order of the \textit{Clarity of Information}.\label{fig:D2}]{
  \includegraphics[width=0.47\textwidth]{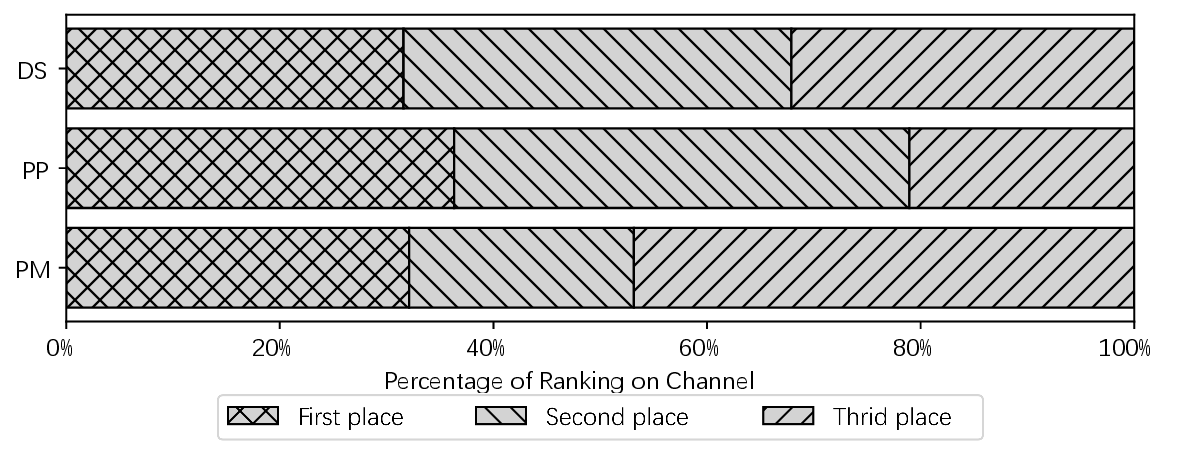}
}\\[6pt]

\subfloat[Rankings in Order of the \textit{Intuitiveness of User Interfaces}.\label{fig:D3}]{
  \includegraphics[width=0.47\textwidth]{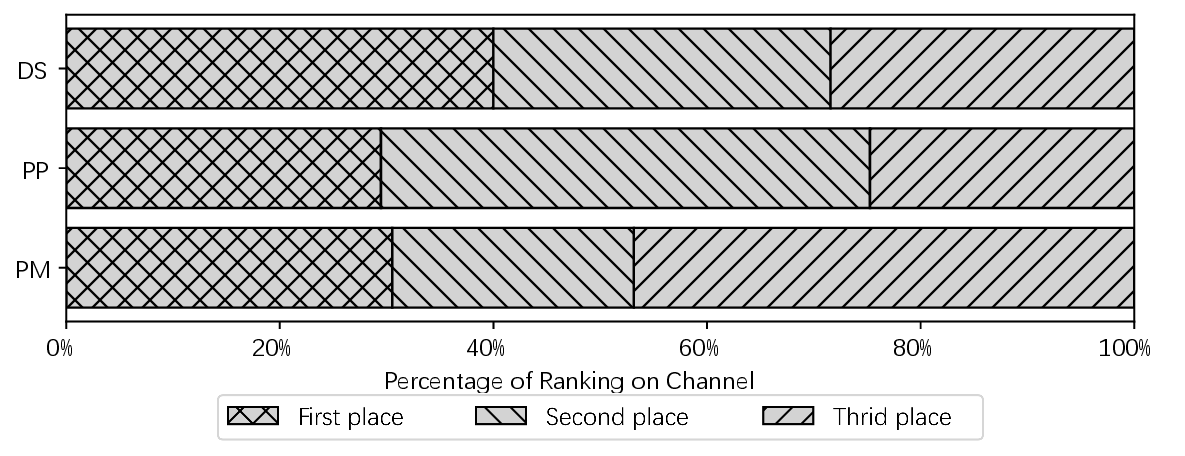}
}\hfill
\subfloat[Rankings in Order of the \textit{Overall Effectiveness}.\label{fig:D4}]{
  \includegraphics[width=0.47\textwidth]{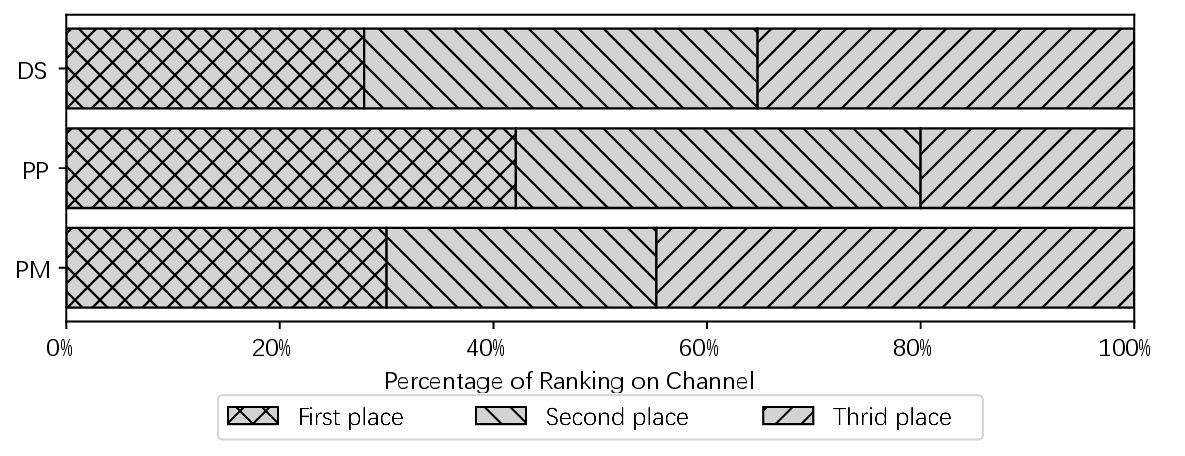}
}

\caption{Participants' Rankings on DS, PP, and PM Channels from four aspects (Questions D.1 to D.4, n=190).}
\label{fig:rankingsD1_to_D4}
\end{figure*}

\textbf{Amount of information.}
As shown in Figure~\ref{fig:D1},
PP is the top channel for both the first place (with 94 or 49.5\% of participants) and first-second places together (with 159 or 83.7\% of participants), 
in terms of the amount of information it provided about the privacy practices of the assigned app (Question D.1).
Among those 159 participants,
47 explained that this channel provides the comprehensive information with in-depth explanations, 
e.g., \textit{``The Privacy Policy had the most information and explained the exact type of data that was being gathered, how it was being used, where it was being stored, how to get it deleted and more''} (USA TODAY~\#8);
17 explained that it provides the detailed disclosure of certain data practices such as collection and use, sharing, or users' data rights,
e.g., \textit{``I like how the privacy policy made it clear that info would not be shared with a 3rd party and the process that should be taken to have your info removed''} (DW~\#41).

\textbf{Clarity of information.}
As for the clarity of information a channel provided about the privacy practices of the assigned app (Question D.2),
PP is again ranked as the top channel for both the first place (with 69 or 36.3\% of participants) and first-second places together (with 150 or 78.9\% of participants) as shown in Figure~\ref{fig:D2}. 
Among those 150 participants, 
37 explained that this channel provides in-depth explanations of privacy practices, e.g., \textit{``I felt the privacy policy laid out each section and had a great explanation for each part and what would be collected''} (USA TODAY~\#18);
two noted its' transparency on privacy practices,
e.g., \textit{``The privacy policy was very transparent and I felt confident that I understood how my information was being used''} (Once~\#9).

\textbf{Intuitiveness of user interfaces.}
As for the intuitiveness of user interfaces (Question D.3),
DS is ranked as the top channel for the first place (with 76 or 40.0\% of participants), while PP is ranked as the top channel for the first-second places together (with 143 or 75.3\% of participants), as shown in Figure~\ref{fig:D3}. 
Among those 143 participants, 
nine explained that a privacy policy is a text document requiring no additional user interactions,
e.g., \textit{``The privacy policy had very little user interface beyond simply scrolling down and reading''} (Bumble~\#42).
Among the 136 or 71.6\% of participants who ranked DS for the first or second place, eight explained that this channel features a layered structure with detailed data collection purposes provided in collapsible components, 
e.g., \textit{``With Data Safety I was appreciative of the process of clicking the arrows to have more info appear''} (Once~\#2).

The detailed results of Questions D.1 to D.3 from the app category, disclosure extensiveness, and channel
sequence aspects are provided in Section 4 of the Supplementary Material~\cite{github_link}.
In summary, for Question D.1, we observed the following statistically significant differences in the first place ranking distributions between the two participant groups: 
30.9\% of dating app participants ranked the DS channel as the first place, compared to only 15.1\% of news app participants; 
63.8\% of S-1 participants ranked the PP channel as the first place, compared to only 35.4\% of S-2 participants; 14.9\% of S-1 participants ranked the PM channel as the first place, compared to 39.6\% of S-2 participants.
For Questions D.2 and D.3, we found no statistically significant differences between the two participant groups 
in any of the three aspects.

\textbf{Overall effectiveness.}
Regarding their overall effectiveness (Question D.4), 
PP is ranked as the top channel for both the first place (with 80 or 42.1\% of participants) and first-second places together (with 152 or 80.0\% of participants) as shown in Figure~\ref{fig:D4}. 
\textbf{\textit{Among those 152 participants}},
45 appreciated this channel's comprehensiveness in minimizing confusion about an app's privacy practices,
e.g., ``\textit{I didn't really understand the privacy and data collection of this app until I read the Privacy Policy. Once I read the Privacy Policy, I knew exactly what data was being collected, why it was being collected, who it was being shared with, how to get it deleted and where it was being stored}'' (USA TODAY~\#8); 
six noted the clear and understandable information provided by it,
e.g., ``\textit{I think the privacy policy was most helpful in helping me understand the privacy practices because it was very detailed with a lot of information that was very easy to follow and understand}'' (Bumble~\#39); 
three mentioned that its information helps them assess privacy risks,
e.g., ``\textit{It was the one that helped me understand the risks the most and also how I could intervene if needed}'' (DW~\#15).
\textit{\textbf{Among the 123 or 64.7\% of participants who ranked DS for the first or second place}}, 11 appreciated its concise and cost-effective information,
e.g., ``\textit{Although privacy policy gave me a lot of information, I feel like data safety gave me the information I wanted to see in a fast and effective matter}'' (USA TODAY~\#10); 
nine mentioned its effective disclosure on data sharing and protection practices,
e.g., ``\textit{Because it provides additional insights into data security measures, which contribute to a better understanding of privacy practices}'' (DW~\#24); 
four noted its information comprehensiveness,
e.g., ``\textit{The data safety tells you exactly what its tracking, how its used and who it may be shared with and why and what rights you have about your data}'' (Bumble~\#4).
\textbf{\textit{Among the 105 or 55.3\% of participants who ranked PM for the first or second place}}, 
18 appreciated its concise and cost-effective information,
e.g., ``\textit{I believe that the permission manifest was very effective in helping me make the decision if I were to want to download this app in the future. The information was clear and concise}'' (USA TODAY~\#9); 
six noted its information comprehensiveness,
e.g., ``\textit{Permission manifest is the most effective because it is clear and comprehensive}'' (USA TODAY~\#5). 

\begin{figure*}[t]
\centering

\subfloat[App Category (D.4)\label{fig:D4-Category_Aspect}]{
  \includegraphics[width=0.32\textwidth]{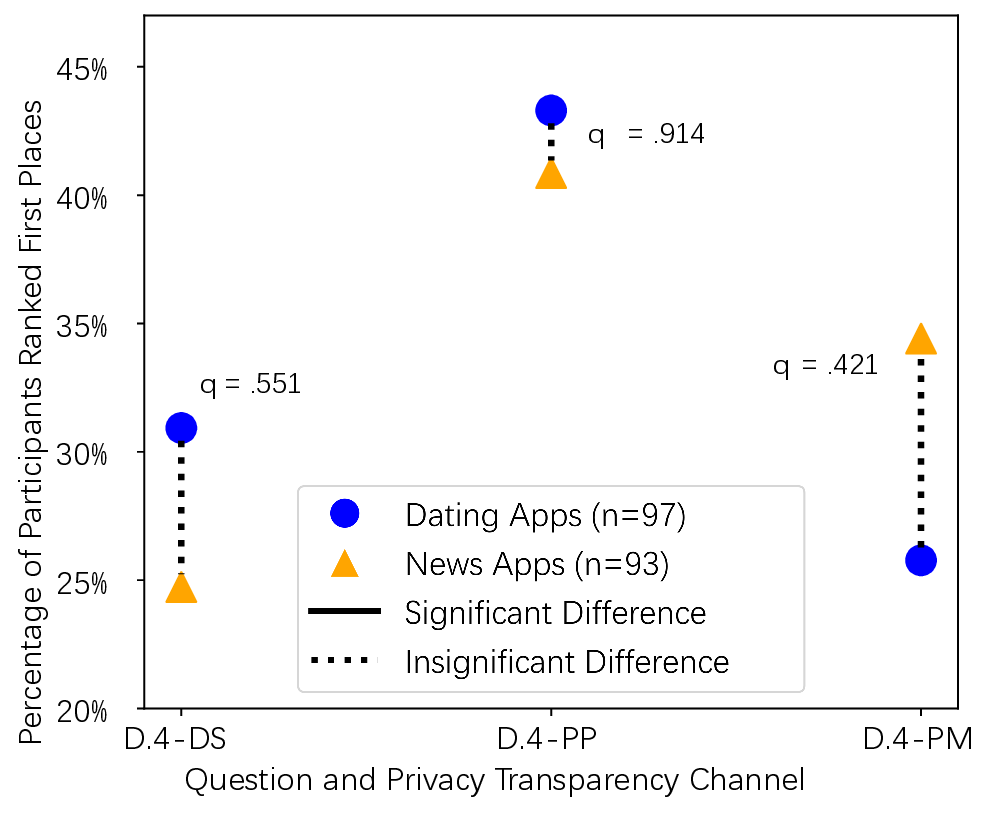}
}\hfill
\subfloat[App Disclosure Extensiveness (D.4)\label{fig:D4-Extensiveness_Aspect}]{
  \includegraphics[width=0.32\textwidth]{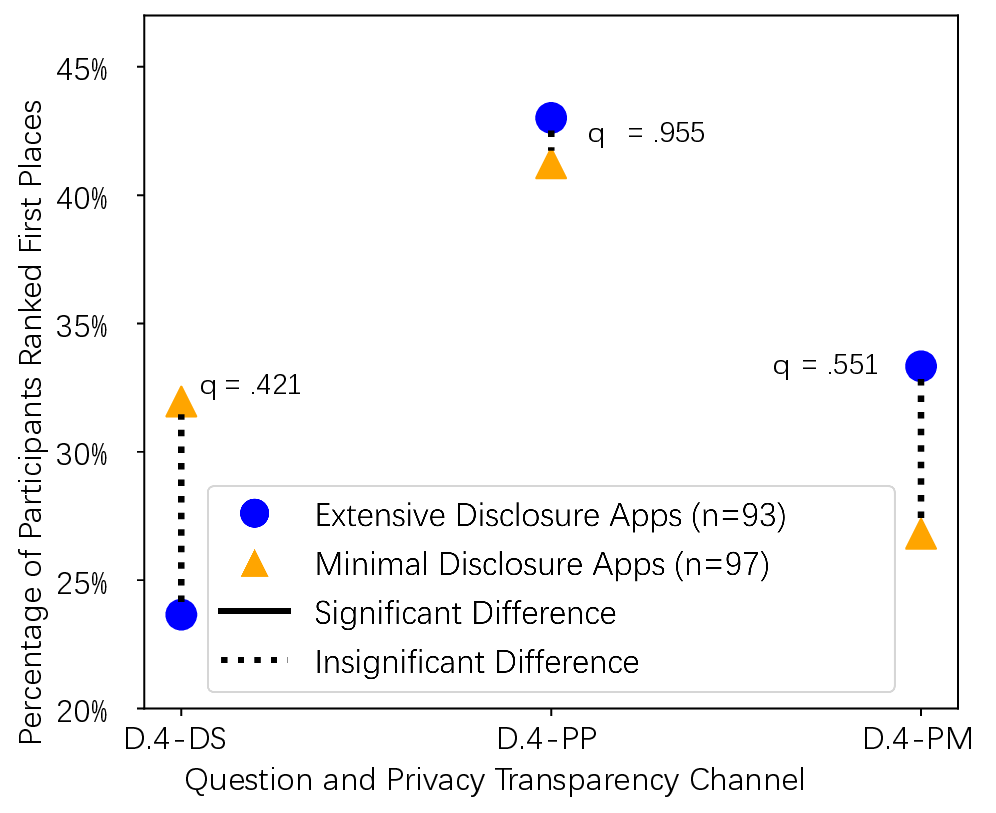}
}\hfill
\subfloat[Channel Sequence (D.4)\label{fig:D4-Sequence_Aspect}]{
  \includegraphics[width=0.32\textwidth]{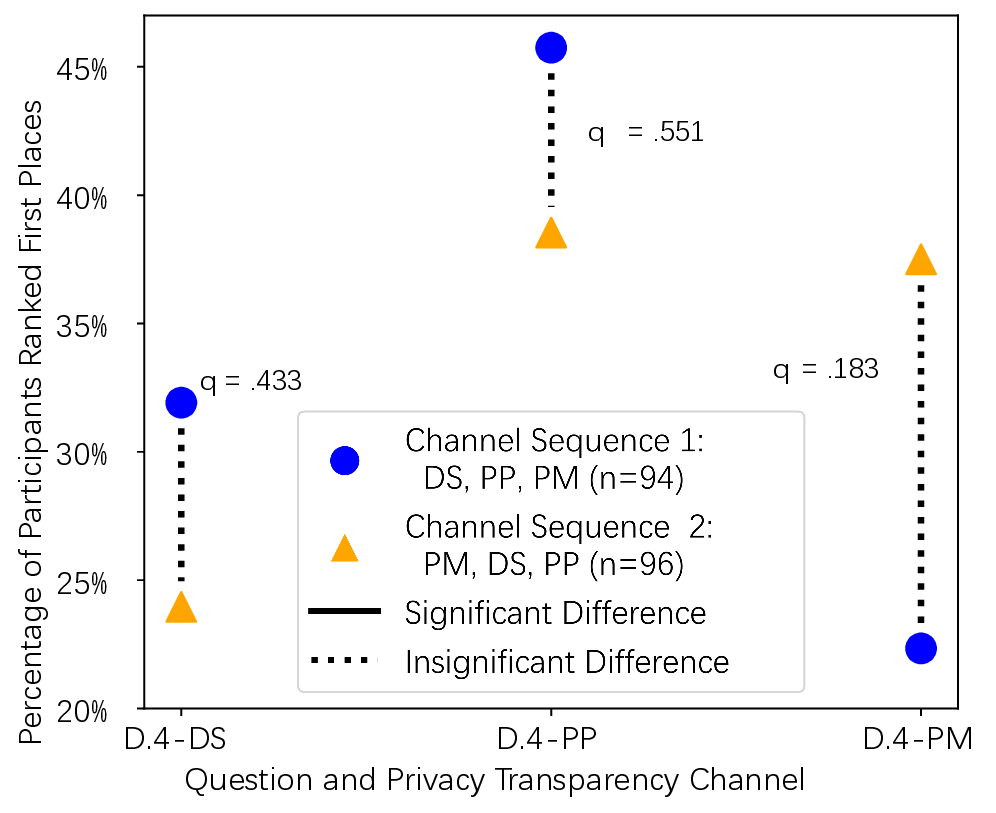}
}

\caption{Details of Participants’ First Place Rankings in Order of the Overall Effectiveness (Question D.4) from Three Aspects.}
\label{fig:D4_From_Three_Aspects}
\end{figure*}

When zooming into details on participants' first place rankings in order of the overall effectiveness (Question D.4), we found \textbf{\textit{no statistically significant differences 
between
the two participant groups in any of the three aspects}}, as shown in  Figures~\ref{fig:D4-Category_Aspect},~\ref{fig:D4-Extensiveness_Aspect}, and~\ref{fig:D4-Sequence_Aspect}.
For each participant group when performing within-subjects comparisons among channels,  
we found that 45.7\% and 22.3\% of S-1 participants ranked the PP and PM channels as the first place, respectively; only this difference is statistically significant~($q=.033$).

We also examined the differences among the three channels, considering all first, second, and third-place rankings.
The Friedman test for Questions D.1 to D.4, respectively, shows that the ranking distributions among the three channels (as shown in Figure~\ref{fig:rankingsD1_to_D4}) \textbf{\textit{differ with statistical significance}}.
For each question, we further performed the post-hoc Nemenyi tests to identify the channel pairs with significant different rankings.
All details are provided in Section~4 of the Supplementary Material~\cite{github_link}.

\textbf{Association analysis.}
We further explored the potential participant associations of the following variables with Question D.4: (1) participants’ overall understanding of an app's privacy practices (Question C.2), (2) their privacy risk concerns (Questions B.7 and C.12), and (3) their privacy sensitivity characteristics (Questions A.1 to A.5).
The Chi-square tests indicate no statistically significant association relationships.
We also analyzed participants' responses to Questions C.2, B.7, C.12, and D.4 per demographic groups (Questions E.1 to E.3) by using the Wilcoxon rank sum test, but found no statistically significant differences in response distributions between participant groups based on the gender, age, and education.
All details are provided in Section~5 of the Supplementary Material~\cite{github_link}.

\textbf{Channel improvement suggestions.}
Multiple-answer Question D.5 asks participants to select the ways that channels can be significantly improved in the future to 
help them better understand the privacy practices of apps and better estimate the privacy risks associated with installing and using apps.
Table~\ref{tab:D5} lists participants' selections of the five provided improvement options with the top-two highlighted.

\begin{table*}[h]
\caption{Participants' Selections on Question D.5 (n=190).}
\scalebox{0.8}{
\begin{tabular}{|p{12.0cm}|p{5.3cm}|}
\hline
\textbf{Channel Improvement Options}     & \textbf{Numbers (and \%) of Participants}  \\ \hline
Provide more and clearer information about an app's privacy practices & \textbf{126~(66.3\%)} \\ \hline
Improve their user interfaces to present an app's privacy practices more effectively & 89~(46.8\%) \\ \hline
Provide a single interface to present their combined information about an app's privacy practices & 101~(53.2\%) \\ \hline
Highlight the potential privacy risks associated with an app's privacy practices & \textbf{139~(73.2\%)} \\ \hline
Make it mandatory for me to interact with them before allowing me to install an app & 70~(36.8\%) \\ \hline
\end{tabular}
}
\label{tab:D5}
\end{table*}

Recall that we have Question C.13 for participants to indicate if a channel they just interacted should be improved in some ways.
As shown in Figure~\ref{fig:C13} in Appendix~\ref{sec:appendix_Additional Results}, 118 (62.1\%), 94 (49.5\%), and 129 (67.9\%) participants agreed that the DS, PP, and PM channels
need improvements, respectively;
the distribution differences between the PP channel and the other two channels, respectively, are statistically significant~($p=.048$ and .041).
\textbf{\textit{Among those 118 participants}}, the top-two 
explanations are: 
16 expected more explanations on privacy practices, 
e.g., ``\textit{I would hope that there would be some plausibly detailed version of the information contained in the transparency channel. It is exceedingly vague}'' (Once~\#5);
15, 14, and nine participants expected more details on data sharing, retention, and use practices, respectively, 
e.g., ``\textit{They should tell you where and for how long your information is stored and who has access to it}'' (USA TODAY~\#11).
\textit{\textbf{Among those 94 participants}}, the top-two 
explanations are: 16 mentioned that privacy policies should be more concise, 
e.g., ``\textit{I guess the info could be a bit more concise and user friendly, but overall it was pretty comprehensive and had a lot of useful information}'' (DW~\#18);
nine expected more transparency on privacy risks,
e.g., ``\textit{The risks of sharing my information under this policy aren't truly explained at all. I'm left to guess what harm might come to me with my information shared this widely. This should definitely be clarified}'' (USA TODAY~\#13).
\textbf{\textit{Among those 129 participants}}, the top-two 
explanations are: 48 expected more explanations on permission
requests,
e.g.,``\textit{It should be stated more clearly what they want these permissions for and what they'll do with them. Are they for basic functionality of the app or expanded features you may not access? Can some more of these be optional?}''~(USA TODAY~\#3); 
19 expected more information on how their data will be used, 
e.g., ``\textit{Information should be given to us on how they intend to use our personal information}''~(Once~\#30).

\textbf{Key RQ3 takeaways or findings.}  
First, PP is considered the most informative (per the amount and clarity of information) and effective channel.
Participants' explanations to their answers to Questions D.1, D.2, and D.4 especially highlighted that information comprehensiveness, in-depth explanations, and minimized confusion about an app’s privacy practices are major advantages of privacy policies.
Second, regarding the first place rankings on channels' overall effectiveness, there are no significant differences between the two participant groups in any of the three aspects.
Third, while the DS and PP channels differ in their presentation styles, they provide more intuitive user interfaces than the PM channel. 
Fourth, three of the five channel improvement options were selected by 
more than half of participants, with ``highlight the potential privacy risks associated with an
app's privacy practices'' being the top choice; meanwhile, 
less than half of participants expected the improvements to the PP channel, but majority of participants expected the improvements to the other two channels.

\section{Discussion}
\label{sec:discussion}

\subsection{Implications and Recommendations}
\label{sec:recommendations}
\textbf{Implications of our key findings.}
Our key findings summarized in Sections~\ref{sec:RQ1 users’ understanding of privacy practices},~\ref{sec:RQ2 users’ judgment of the privacy risks}, and~\ref{subsec:RQ3_overall_opinions} imply that 
(1) all three channels are effective in increasing participants' overall understanding of an app's privacy practices, (2) the data safety and permission manifest channels are particularly effective in raising participants' concerns about the overall privacy risks of an app, and (3) each channel has its own pros and cons.

We found that DS is regarded as the channel with the most intuitive user interfaces; meanwhile, it helps participants more accurately capture apps' data collection and use as well as data sharing practices, compared to the other two channels.
On the other hand, its overall effectiveness is outperformed by the PP channel, as many participants valued the latter’s comprehensiveness in reducing confusion about an app’s privacy practices.
Moreover, we identified shortcomings in the DS channel's disclosure of data use purposes, data rights, and data storage or retention practices, indicating that those are some potential areas for improvement.

We found that PP is 
regarded as the most informative and effective channel. Through this channel with its rich information and text document format, participants are more likely to perceive some online or physical risks.
Nevertheless, likely due to its' 
lengthy nature, only a small percentage
of participants accurately captured the data rights and data storage or retention related information that is largely only provided by this channel.

We found that PM is the least informative and effective channel among the three.
However, this channel excels at raising participants' concerns about the overall privacy risks of an app. 
This may be due to the unique information presented in it that reflects an app's actual practices, for example, participants might not find hardware access related information in other channels and could be surprised by a list of such permission requests; another possible reason could that some participants 
may not understand the Android runtime permission model~\cite{AndroidRuntimePermissions} and may consider that all listed permissions need to be granted.

Based on the zoomed-in between-subjects analyses, we found some statistically significant differences only from the app category and disclosure extensiveness aspects in terms of participants' overall understandings of an app's privacy practices and concerns about
the overall privacy risks, as summarized in the key takeaways for RQ1 and RQ2.
Regarding the first place rankings on channels’ overall
effectiveness, there are no significant differences between the two participant groups in any of the three aspects, as summarized in the key takeaways for RQ3.

Comparing our key findings and implications with previous work is difficult because no study has compared the effectiveness of these channels. 
It is also hard to compare our results on individual channels with previous work due to the study design and focus differences; for example, neither users' privacy risk concerns nor changes in their perceptions were analyzed in privacy label studies such as~\cite{lin2023data, zhang2022usable}.
Thus, we can at most connect some of our specific results with previous work.
Especially, on the shortcomings of the DS channel in disclosing data use purposes, our results echo those in~\cite{lin2023data, zhang2022usable}.
Beyond confirming the PP channel's issues such as users' struggles with technical terms as in~\cite{gluck2016short, chen2023investigating, tang2021defining, ibdah2021should}, we further found that it remains the most informative and effective channel. 
Prior studies (e.g.,~\cite{cao2021large}) revealed that users do not have the context for the declared permissions before installing an app, similar to what we observed in Section~\ref{subsec:impact of channels on RQ2}; we further found that permission 
manifests obviously raised users’ privacy risk concerns.

Note that although these three channels provide useful privacy transparency information, they remain insufficient to fully inform users about potential privacy risks or to prevent privacy violations.
The accuracy and completeness of the information disclosed through these channels depend on developers’ attention and integrity.  Users should be cautious that inconsistencies and even deliberately misleading disclosures may exist.
In addition, extracting privacy-related comments from app reviews (e.g., using natural language processing techniques~\cite{hatamian2019revealing, nema2022analyzing}) may offer users with complementary information on apps’ privacy practices.
\red{Also note that beyond privacy transparency information, users’ privacy decisions can also be shaped by cognitive and behavioral factors~\cite{wisniewski2020predicting, fleming2023tell}.}

\textbf{Recommendations for the Google Play Store and similar app store providers to improve the design of their privacy transparency channels.} 
First, 
the DS channel may require app developers to provide more types of information (e.g., data rights and data storage or retention practices) to increase its disclosure comprehensiveness, help reduce users' confusion about privacy practices, and boost its overall effectiveness.
Second, 
it can be helpful to incorporate short in-line explanations (as commented by 16 participants to their answers to Question C.13) or tooltips-based examples to clarify some important data use and sharing purposes directly in the DS channel, instead of only providing a general link~\cite{UnderstandPlayStoreDataSafety} and expecting users to further read the details. 
Third, the PM channel could be improved by concisely incorporating contextual information (e.g., when and why certain permissions are needed), explaining technical terms, reminding users about the Android runtime permission model, and clarifying the differences between this channel and the data safety channel. 
Fourth, while being 
most informative and effective, PP is the least often checked channel (by only 43.7\% of participants as shown in Section~\ref{subsec:Demographics_Privacy-Sensitivity} from the responses to Question C.1); thus, 
it can be necessary and helpful to make the privacy policy URL more accessible to users by potentially relocating it to an app's data safety summary section on the main page, instead of keeping it at the end of an app's detailed data safety page.

\textbf{Recommendations for developers to improve their apps' privacy transparency in channels.} 
First, developers should be careful when filling out an app's data safety form, ensuring not only the completeness of the disclosed privacy practices but also the consistency with the information provided in the app's privacy policy and permission manifest. 
Specifically, app developers may leverage new techniques (such as an IDE plugin~\cite{li2024matcha}) to automatically generate accurate data safety labels.
Second, although many participants appreciated the comprehensiveness of privacy policies in helping them reduce confusion and better estimate privacy risks, the terminology complexity and length of those policies have limited their utility. 
Summarizing and condensing privacy policies are still active and important research topics, with many techniques being proposed over the years (e.g.,~\cite{gluck2016short, wilson2016crowdsourcing}).
App developers may leverage such techniques to make their privacy policies more concise (with key privacy practices summarized at the beginning) and easier to understand (by using consistent privacy practice section headings).
In addition, app developers may consider using interactive designs such as drag and drop actions to actively engage users with privacy policy content and improve their attention to key information~\cite{karegar2020dilemma}.
Third, four participants explained their PP channel improvement answers (to Question C.13) that it was unclear if they are afforded the same rights (e.g., data rights or controls typically stated in regulation-specific sections) as users who live in GDPR or CCPA protected countries or regions.
Developers may clarify these in apps' privacy policies and provide 
relevant information to users in other countries or regions too.

\textbf{Recommendations for app users to consider in their future app selection and installation.} 
First, users should learn as much information as possible from all three channels before installing an app, 
as they offer unique perspectives into apps' privacy practices and can collectively help users better assess privacy risks. 
Second, 
users should not underestimate the privacy risks of those apps (e.g., news apps) in the ``low'' privacy sensitivity level categories, and should carefully check their privacy transparency information to make informed decisions.

\vspace{-1pt}
\subsection{Limitations and Future Work}
\label{sec:limitation_and_future_work}
\textbf{The study design perspective.}
We considered two randomized channel interaction sequences that reflect the actual channel access design on the Google Play Store, in which the data safety channel is always placed before (thus potentially primes) the privacy policy channel.
Thus, the latter's better performance over the former 
on informativeness and overall effectiveness could partially be attributed to this potential priming effect.
In our design, we aimed to mitigate the internal validity concerns (due to order and carry-over effects) by (1) separating the webpages of the three channels clearly, and (2) asking participants to answer our during-interaction questions purely based on what they learned from a current channel.
However, a participant's responses to a channel that appears later in a sequence could still be affected by a channel that appears early.
In addition, as detailed in Section~\ref{sec:Simulated User-Channel Interactions}, we extracted the key content from each original privacy policy and presented 
the condensed content in the PP channel mainly due to the limited time frame of the study.
Participants may view our condensed privacy policies more favorably than the original ones, which could potentially increase their effectiveness.
Future work may address this limitation by evaluating the effectiveness of full-length privacy policies during users’ real-world app installation processes.
In addition, we used participants' responses per app descriptions as a baseline for comparison.
While this design aligns with how users typically review apps during real-world app discovery, future work may investigate the absolute effectiveness of the channels without providing app descriptions.

\textbf{The results perspective.}
We derived the effectiveness results of the three channels through a semi-controlled study in which participants spent, on average, 11.9 (SD=8.1), 12.5 (SD=8.9), and 9.0 (SD=5.8) minutes interacting with the DS, PP, and PM channels, respectively. 
However, in the real-life app selection process, many users may not check these channels at all, or do not often check them (as shown in Section~\ref{subsec:Demographics_Privacy-Sensitivity} from the responses to Question C.1).
Moreover, even among users who do check some of these channels, many may only briefly read the information before installing an app.
Thus, the channel effectiveness results reported in our study might only represent an upper bound obtained in a lab setting.
In addition, less than 10\% of our participants had used the four apps before (Section~\ref{subsec:Demographics_Privacy-Sensitivity}) \red{and 97.4\% of participants used a browser on a PC or a Mac to complete the study}.
While challenging, it can be worthwhile for future studies to investigate these channels' effectiveness in real-life settings and recruit more participants who had used the apps before.
Future studies may also track users' fine-grained interactions with each channel (along with the timing information) to potentially identify additional insights beyond purely analyzing their responses to questions.
A higher hourly compensation rate is also suggested for future studies.

\textbf{The privacy regulation perspective.}
We selected apps only from the U.S. Google Play Store and recruited participants residing in the U.S., where no federal-level across-sector privacy regulation has been enacted. 
Therefore, questions from the privacy regulation perspective cannot be answered by our study. 
For example, it is unclear whether users living in countries or regions under the jurisdiction of privacy regulations (e.g., GDPR) have different opinions about the three channels, which could be an interesting topic for future research.

\section{Conclusion}
\label{sec:conclusion}
We conducted an interaction study to investigate the effectiveness of the three privacy transparency channels provided by the Google Play Store.  
We found that (1) all
three channels are effective in increasing
participants’ overall understanding of an app’s privacy
practices, 
(2) the data safety and permission manifest channels are particularly effective in raising participants’ concerns
about the overall privacy risks of an app, and (3) each channel has its own pros and cons.
These channels complement each other and should all be improved.
We provided recommendations
to stakeholders for achieving better privacy transparency.
Our study offers a novel comparative perspective on highlighting the three channels' individual and complementary roles in privacy transparency; it prompts researchers to further consider 
how different forms of information and user interface designs may collectively benefit users in digital environments.







\bibliographystyle{unsrturl}
\bibliography{privacy_channel}

\begin{appendices}
    \section{Methodological Transparency and Reproducibility}
\label{sec:appendix_META}

Appendix~\ref{sec:appendix_five_sets_of_ questions} provides the complete five sets of questions
designed and used in this study. We attached the source code of our web app in the supplementary file ``web\_app\_code.zip'' for reviewers to inspect.
We also created a GitHub link (Reference~\cite{github_link}) to the source code of our web app, and cited it in Section~\ref{sec:introduction}.

\section{Some Screenshots}
\label{sec:appendix_screenshots}
\begin{figure*}[t]
\centering

\subfloat[data safety summary section\label{fig:data_safety_summary_example}]{
  \includegraphics[width=0.245\textwidth]{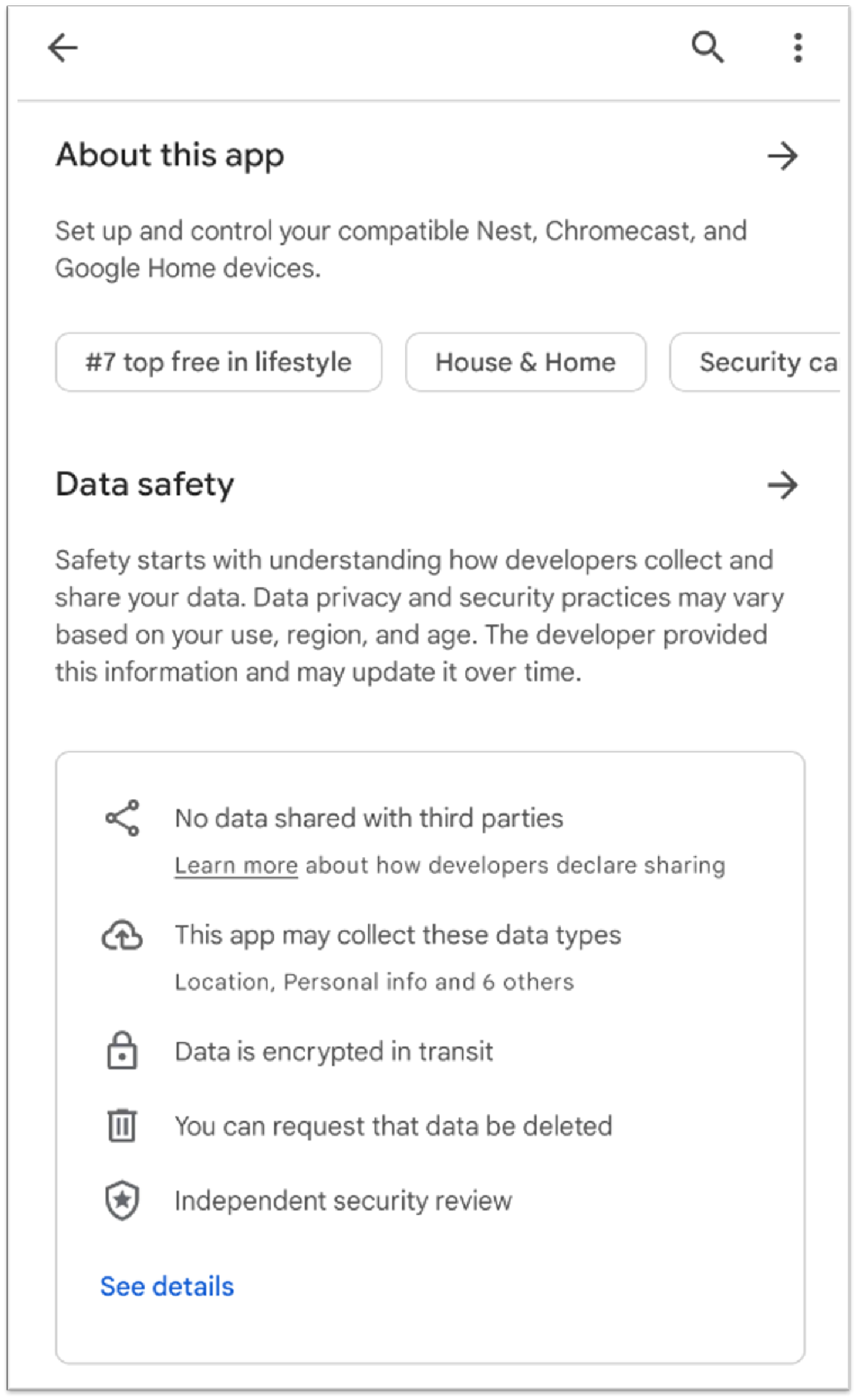}
}
\subfloat[detailed data safety page\label{fig:data_safety_example}]{
  \includegraphics[width=0.245\textwidth]{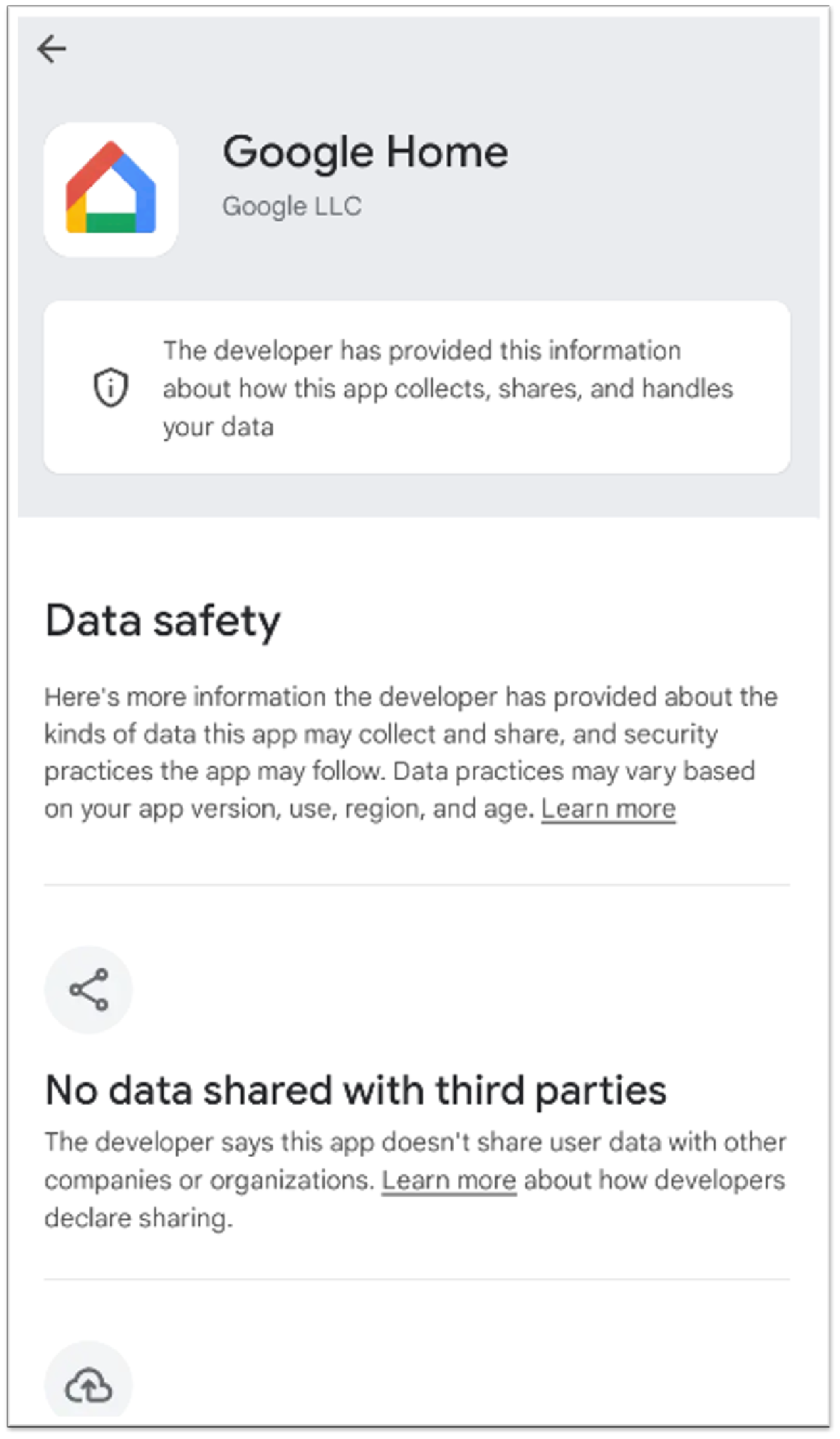}
}
\subfloat[privacy policy\label{fig:privacy_policy_example}]{
  \includegraphics[width=0.245\textwidth]{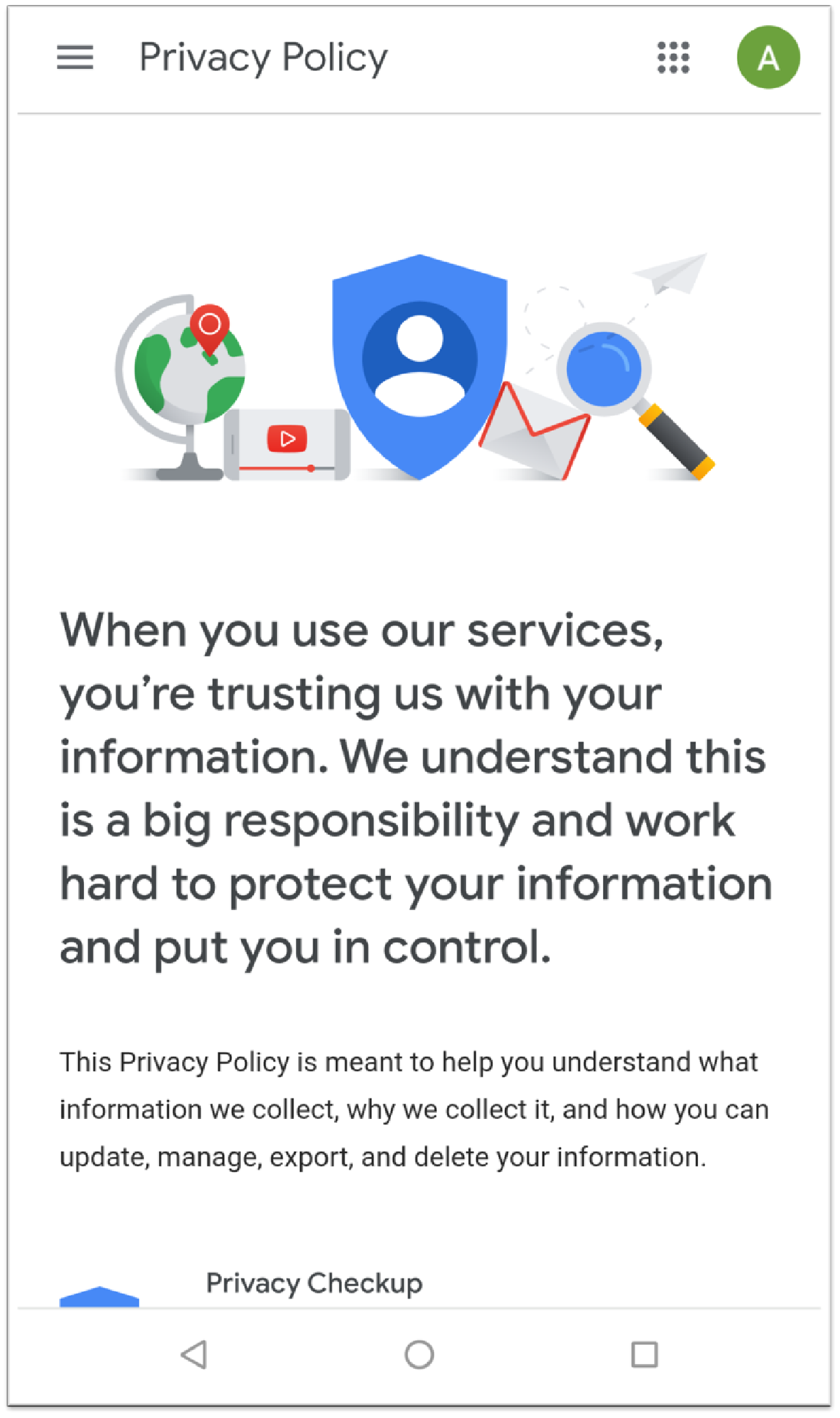}
}
\subfloat[permission manifest\label{fig:permission_example}]{
  \includegraphics[width=0.245\textwidth]{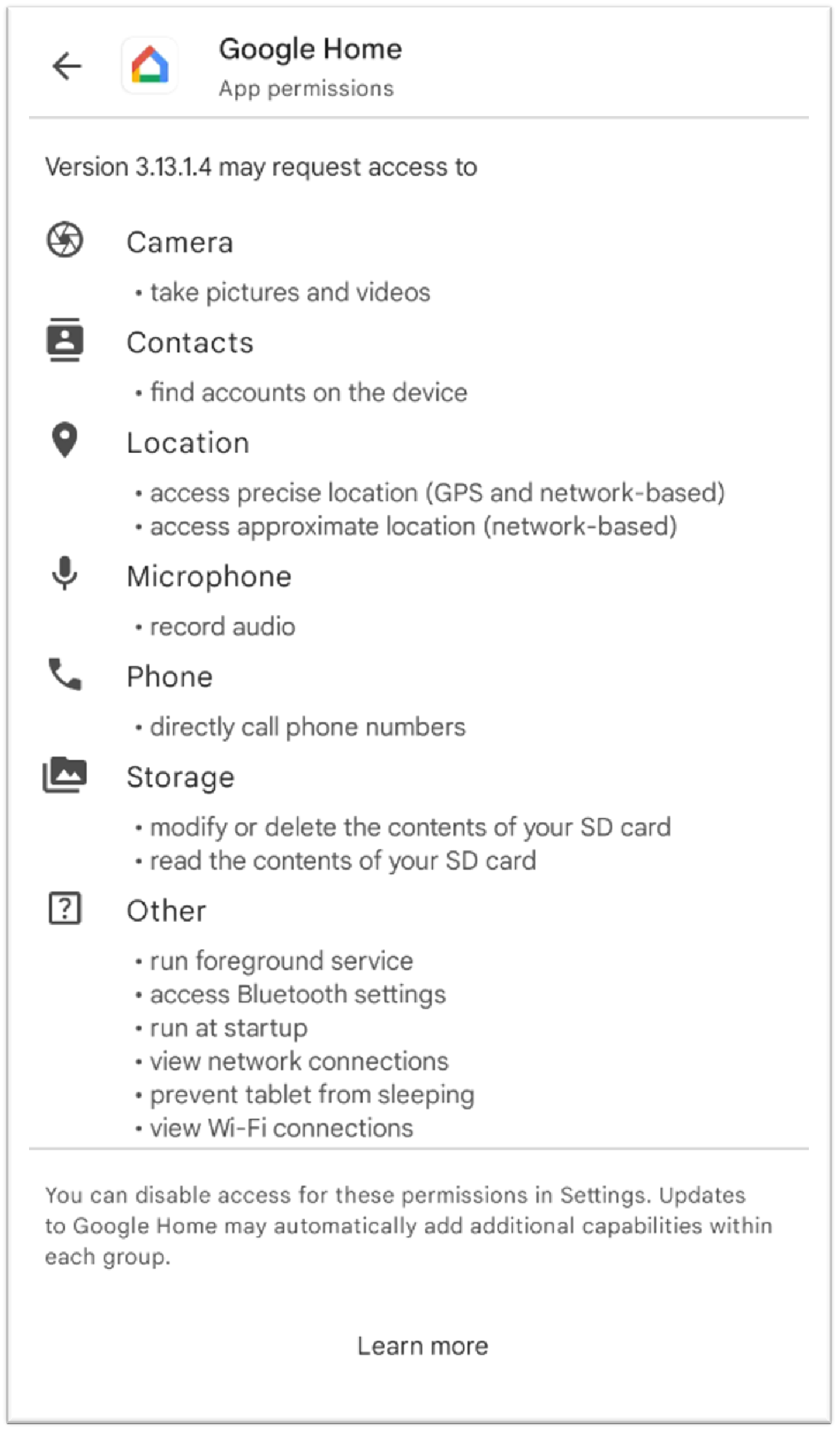}
}

\caption{Privacy Transparency Channels for the Google Home App. Sub-figures (a) and (b) are the interfaces for the data safety channel; (c) and (d) are the interfaces for the privacy policy and permission manifest channels, respectively.}
\label{fig:channels_examples}
\end{figure*}

\begin{figure*}[t]
\centering

\subfloat[App's main page\label{fig:interface_app_A}]{
  \includegraphics[width=0.38\textwidth]{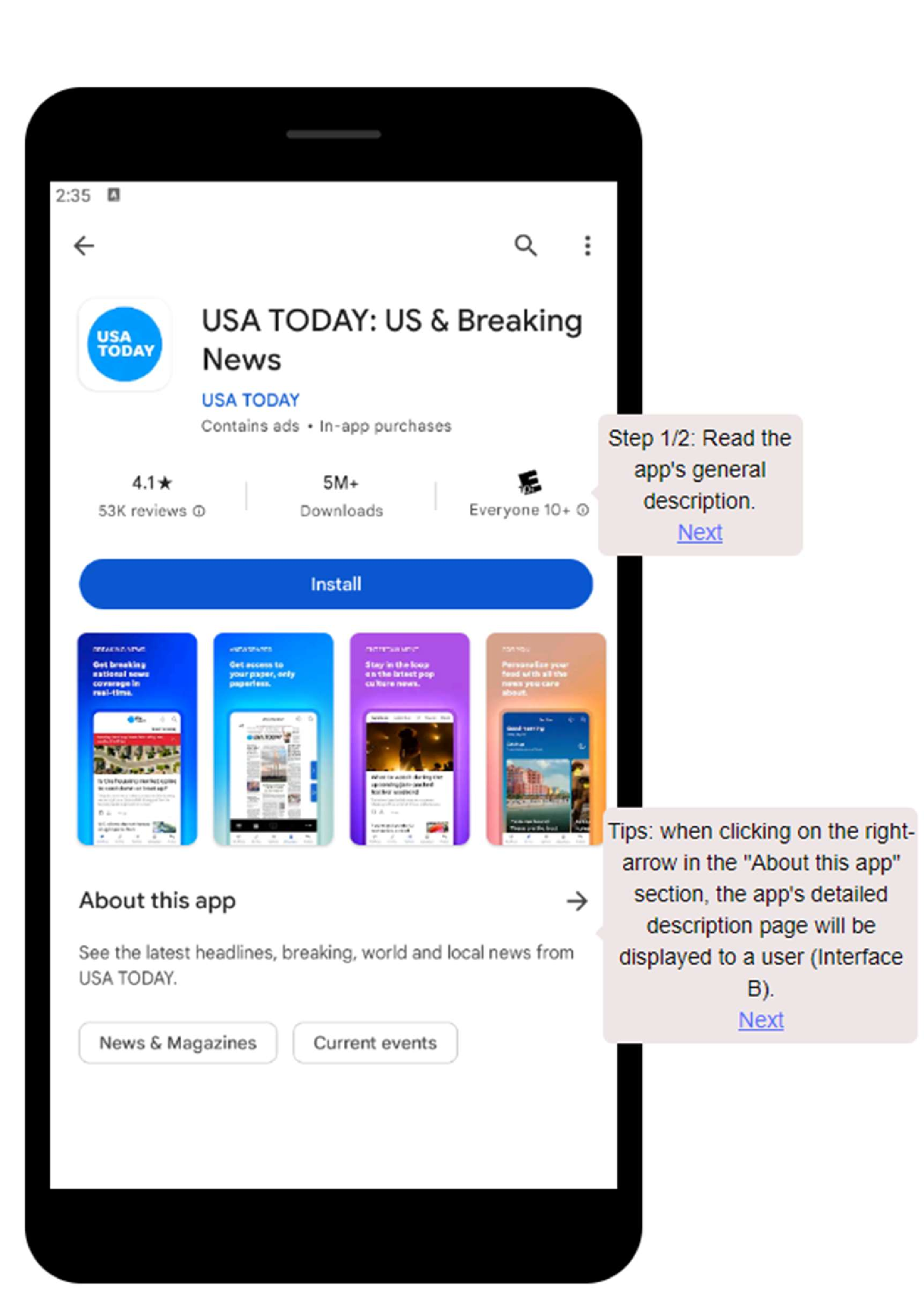}
}
\subfloat[App's detailed description page\label{fig:interface_app_B}]{
  \includegraphics[width=0.36\textwidth]{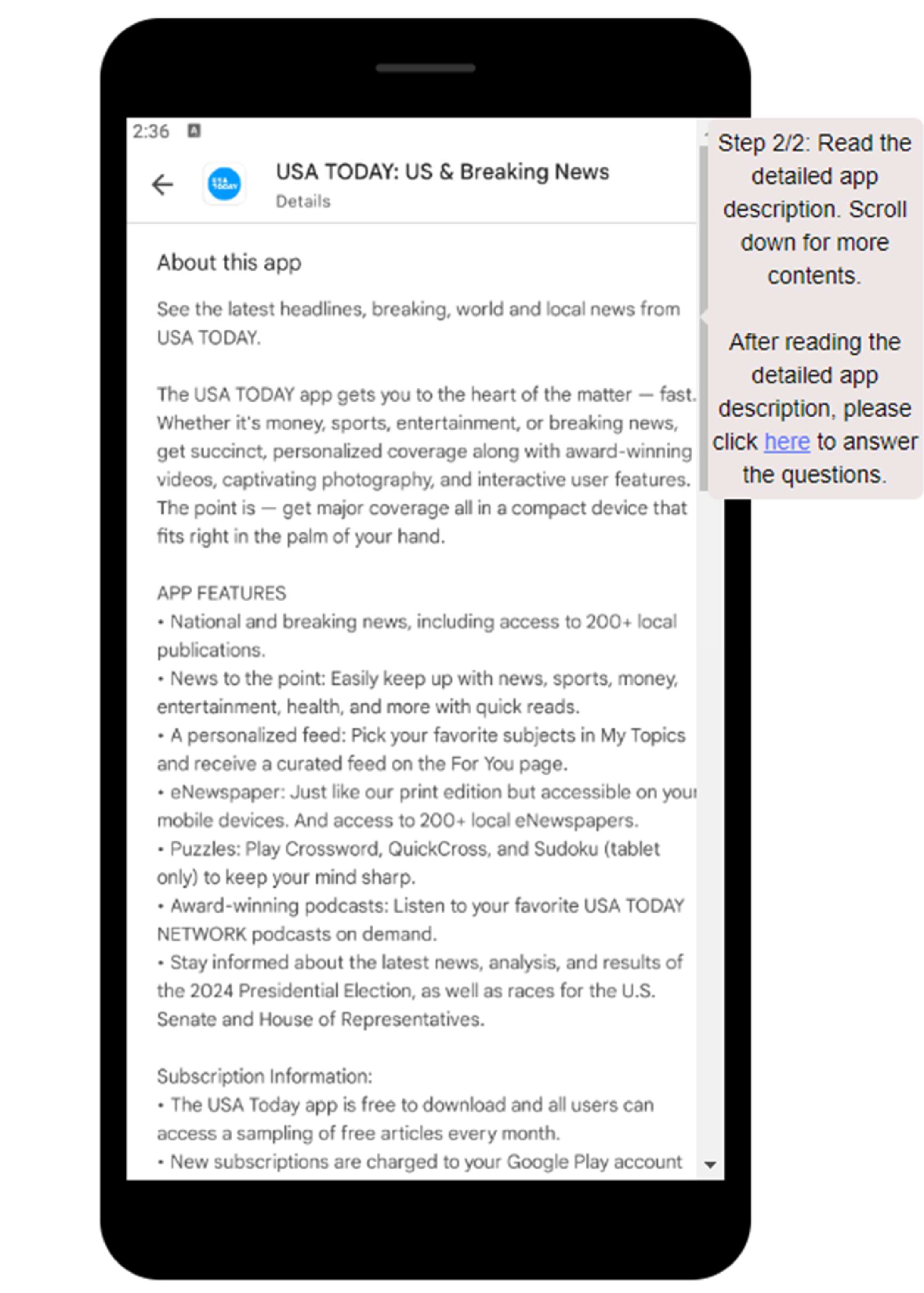}
}

\caption{Example of Our Simulated Main Page and Detailed Description Page of the USA TODAY App.}
\label{fig:interface_app}
\end{figure*}

\begin{figure*}[t]
\centering

\subfloat[App's data safety summary section on the main page\label{fig:interface_data_safety_A}]{
  \includegraphics[width=0.4\textwidth]{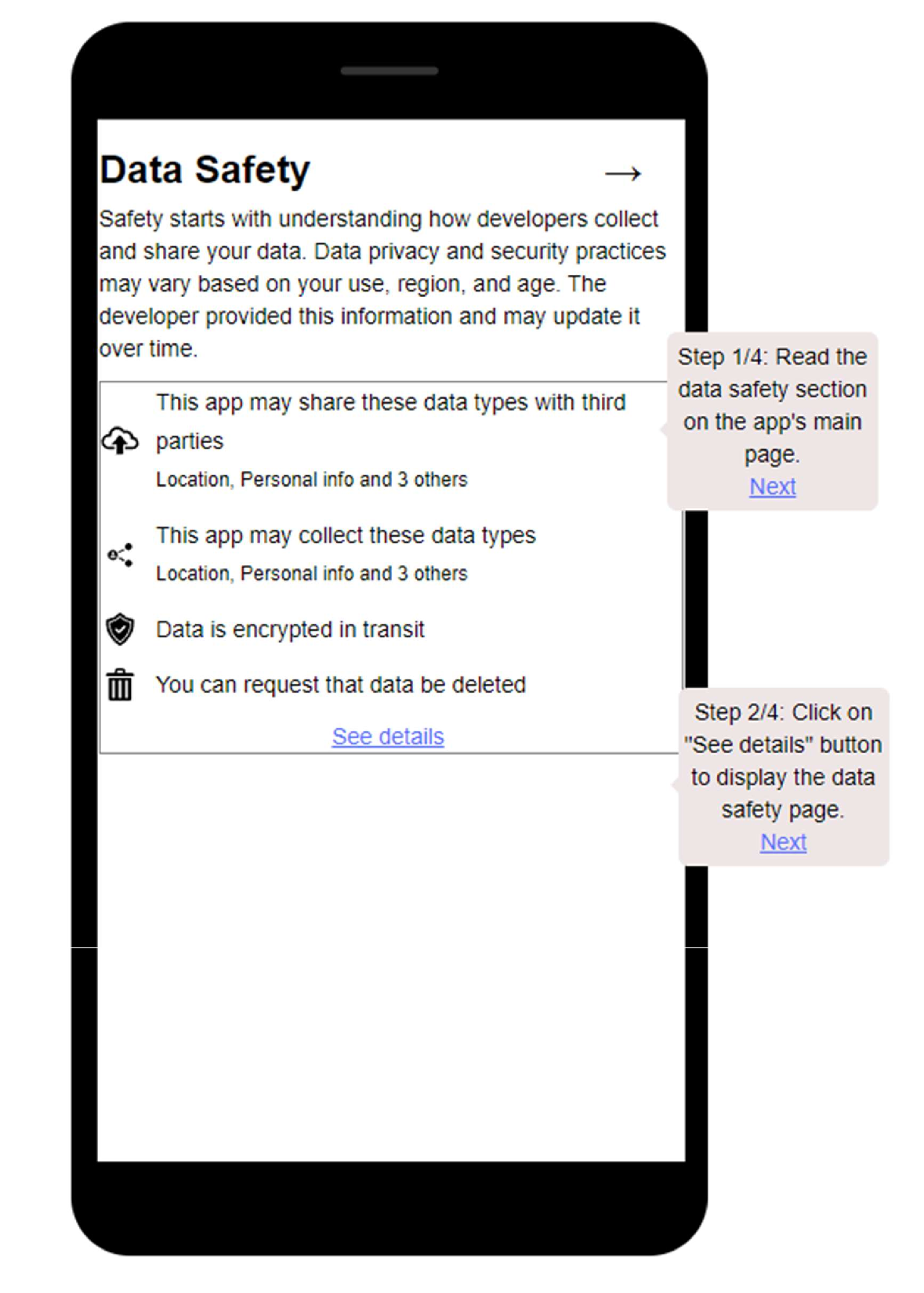}
}
\subfloat[App's detailed data safety page\label{fig:interface_data_safety_B}]{
  \includegraphics[width=0.4\textwidth]{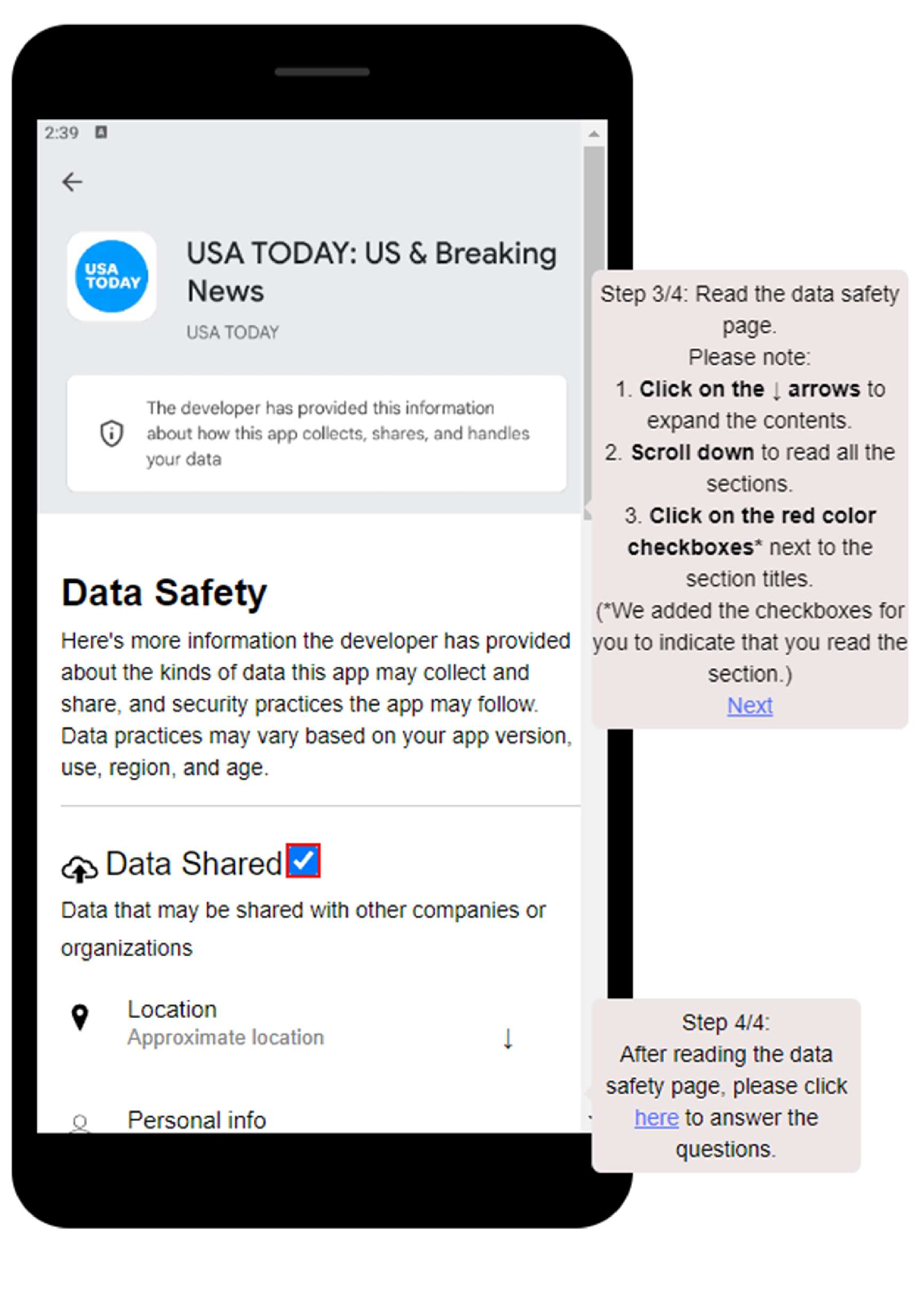}
}

\caption{Example of Our Simulated Data Safety Channel Interfaces of the USA TODAY App.}
\label{fig:interface_data_safety}
\end{figure*}

\begin{figure}[h!]
  \centering
  \includegraphics[width=0.6\linewidth]{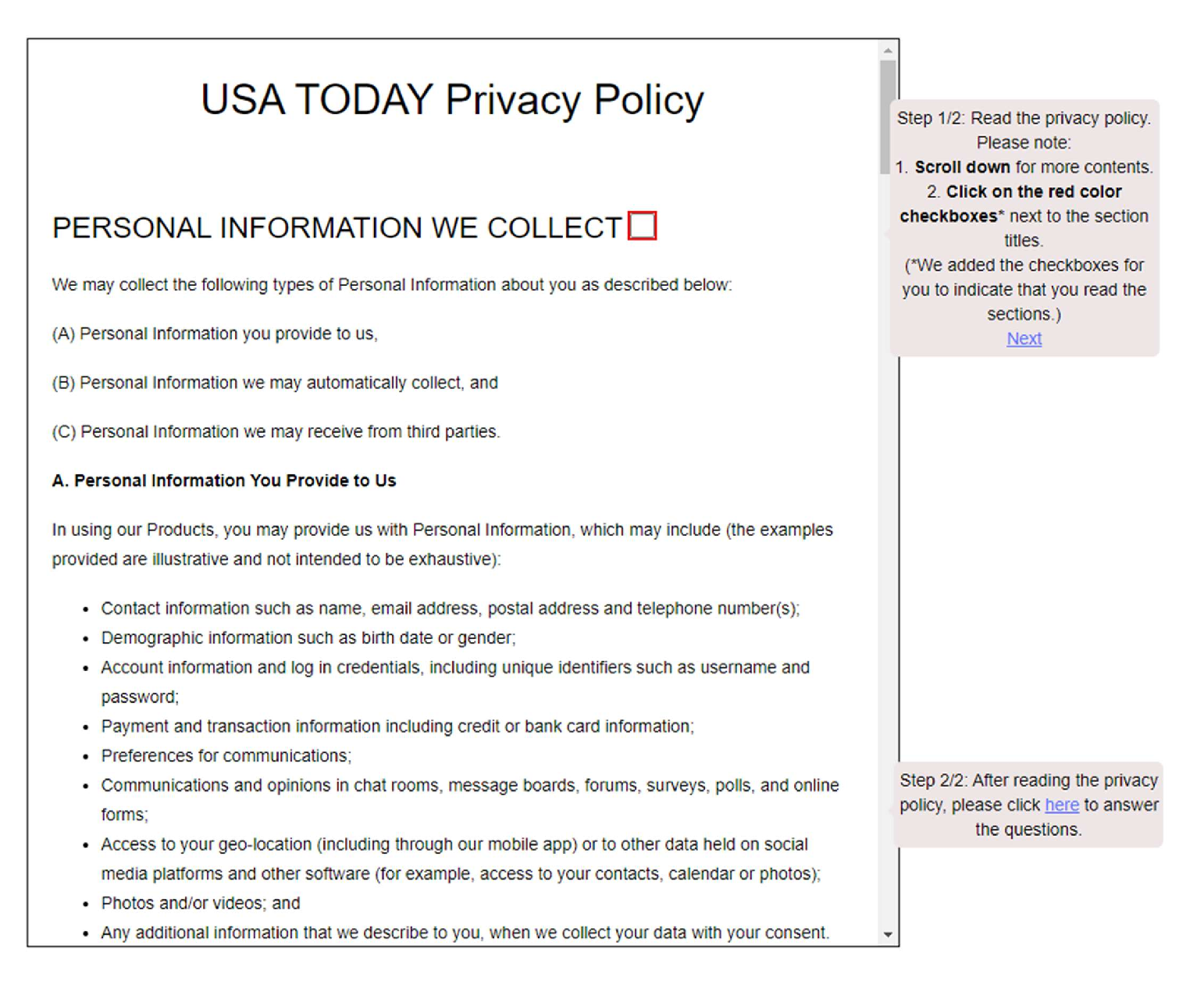}
  \caption{Example of Our Simulated Privacy Policy Channel Interface of the USA TODAY Mobile App.}
  \label{fig:interface_privacy_policy}
  \vspace{-15pt}
\end{figure}

\begin{figure}[h!]
  \centering
  \includegraphics[width=0.4\linewidth]{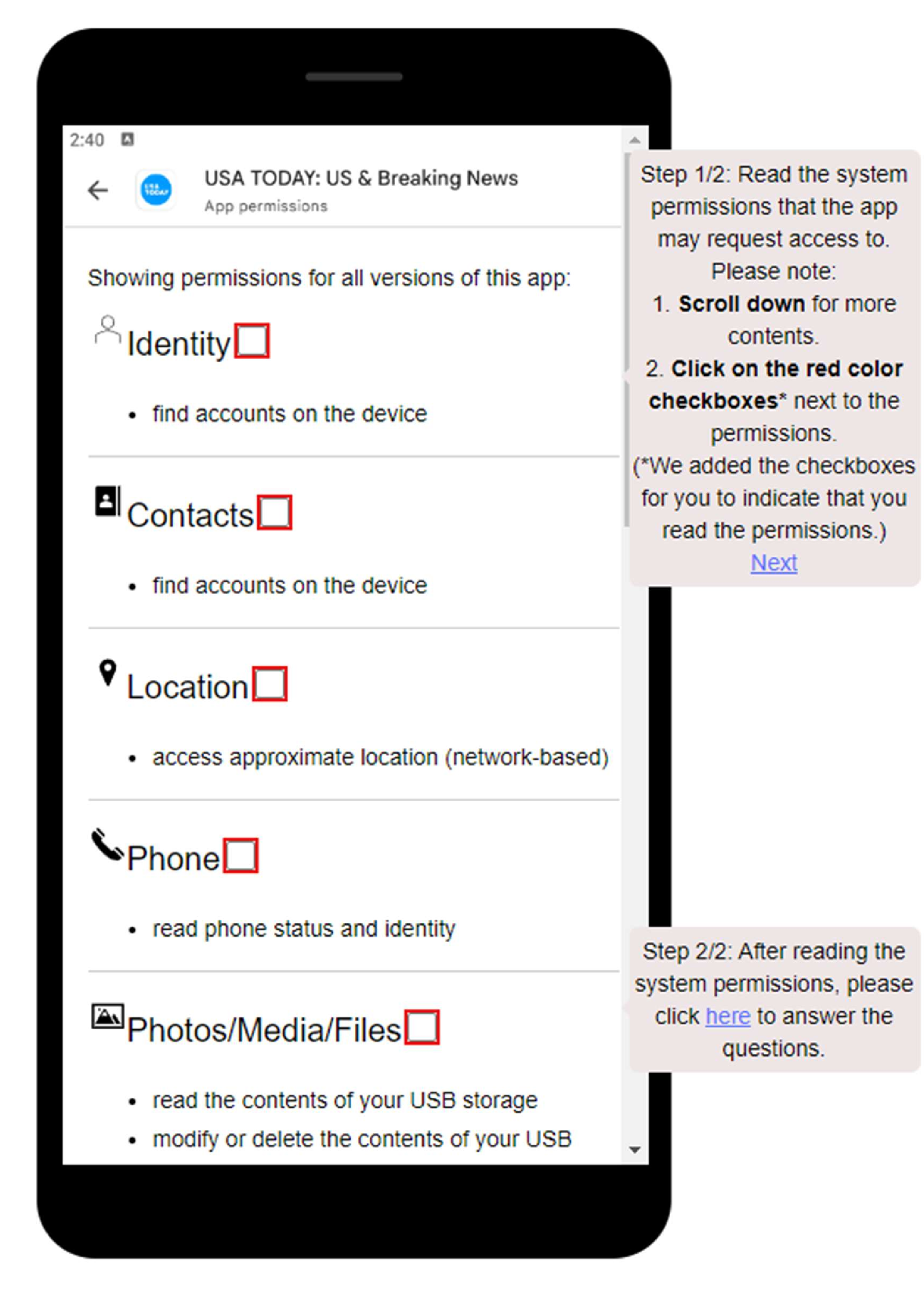}
  \caption{Example of Our Simulated Permission Manifest Channel Interface of the USA TODAY Mobile App.}
  \label{fig:interface_permission_manifest}
  \vspace{-15pt}
\end{figure}

Figure~\ref{fig:channels_examples} shows 
the transparency channels for the Google Home app. 
Figures~\ref{fig:interface_app}~to~\ref{fig:interface_permission_manifest} 
show examples of our simulated interfaces of the USA TODAY app.

\section{The Five Sets of Questions}
\label{sec:appendix_five_sets_of_ questions}
\subsection{Pre-Study Questions}
\label{sec:appendix_Pre-Study Questions}

\noindent\textbf{A.1} Based on my experience of using mobile apps, I consider the following types of information privacy-sensitive (Please select all that apply; if no option is applicable, please select “None of above”.):

\noindent\Hsquare~Location
\Hsquare~Personal info (e.g., name, email address, etc.)
\Hsquare~Financial info (e.g., purchase history, payment info, etc.)
\Hsquare~Health and fitness info
\Hsquare~Messages (e.g., emails, SMS)
\Hsquare~Photos and videos
\Hsquare~Audios
\Hsquare~Files and docs stored on the device
\Hsquare~Calendar events
\Hsquare~Contact list
\Hsquare~App activity (e.g., in-app search history and interaction, etc.)
\Hsquare~Web browsing history
\Hsquare~App info and performance (e.g., app crash log)
\Hsquare~Device or other identifiers (e.g., phone number, etc.)
\Hsquare~Others (please specify)
\Hsquare~None of above

\noindent \textbf{A.2} Some of my mobile apps collect, use, and potentially share (with other parties) some types of privacy-sensitive information that I selected above in Question A.1.

\noindent$\circ$ Strongly agree 
$\circ$ Somewhat agree
$\circ$ Neither agree nor disagree
$\circ$ Somewhat disagree
$\circ$ Strongly disagree

\noindent \textbf{A.3} I or someone I know had an experience of being a victim of some mobile app related privacy violation.

\noindent $\circ$ Yes $\circ$ No $\circ$ Unsure 

\noindent \textbf{A.4} It is important for me to know how a mobile app collects, uses, and potentially shares (with other parties) my privacy-sensitive information before deciding to install it.

\noindent $\circ$ Strongly agree ... $\circ$ Strongly disagree  

\noindent \textbf{A.5} In general, I am concerned about the overall privacy risks associated with installing and using mobile apps.

\noindent $\circ$ Strongly agree ... $\circ$ Strongly disagree

\subsection{Pre-Interaction Questions}
\label{sec:appendix_Pre-Interaction Questions}

\noindent \textbf{B.1} I installed and used the [mobile app] on my smartphone before.

\noindent $\circ$ Yes $\circ$ No $\circ$ Unsure 

\noindent \textbf{B.2} Based on what I learned so far from the app description, my privacy-sensitive information could be collected, used, and potentially shared (with other parties) by the [mobile app] (if I install and use it).

\noindent $\circ$ Strongly agree ... $\circ$ Strongly disagree 

\noindent \textbf{B.3} Based on what I learned so far from the app description, the collection, use, and potentially sharing (with other parties) of my privacy-sensitive information by the [mobile app] (if I install and use it) may harm me by incurring some online risks such as restraining me from expressing my views (e.g., political views) free from worry, allowing attackers to steal my identity, and/or decreasing my chances of getting a good job in the future.

\noindent $\circ$ Strongly agree ... $\circ$ Strongly disagree 

\noindent \textbf{B.4} Based on what I learned so far from the app description, the collection, use, and potentially sharing (with other parties) of my privacy-sensitive information by the [mobile app] (if I install and use it) may harm me by incurring some physical risks (e.g., stalking and burglary) to me or my family.

\noindent $\circ$ Strongly agree ... $\circ$ Strongly disagree 

\noindent \textbf{B.5} Based on what I learned so far from the app description, the collection, use, and potentially sharing (with other parties) of my privacy-sensitive information by the [mobile app] (if I install and use it) may benefit me by providing me with the customized service tailored to my preferences or needs.

\noindent $\circ$ Strongly agree ... $\circ$ Strongly disagree 

\noindent \textbf{B.6} Based on what I learned so far from the app description, the collection, use, and potentially sharing (with other parties) of my privacy-sensitive information by the [mobile app] (if I install and use it) may benefit me by providing me with the enhanced online security protection such as better fraud detection or prevention.

\noindent $\circ$ Strongly agree ... $\circ$ Strongly disagree 

\noindent \textbf{B.7} Based on what I learned so far from the app description, I am concerned with the overall privacy risks associated with installing and using the [mobile app].

\noindent $\circ$ Strongly agree ... $\circ$ Strongly disagree 

\noindent Please explain your answer to Question B.7 with at least 15 words and with high quality.

\subsection{During-Interaction Questions}
\label{sec:appendix_During-Interaction Questions}

\noindent \textbf{C.1} In my previous experiences with installing mobile apps from the Google Play Store, I often reviewed the privacy practice information disclosed on the [transparency channel] before deciding to install a mobile app.

\noindent $\circ$ Strongly agree ... $\circ$ Strongly disagree

\noindent \textbf{C.2} Based on the privacy practices I learned purely from the [transparency channel], my privacy-sensitive information could be collected, used, and potentially shared (with other parties) by the [mobile app] (if I install and use it).

\noindent $\circ$ Strongly agree ... $\circ$ Strongly disagree

\noindent \textbf{C.3} Based on the privacy practices I learned purely from the [transparency channel], the [mobile app] (if I install and use it) may collect and use the following types of privacy-sensitive information (Please select all that apply; if no option is applicable, please select “None of above”.):

\noindent\Hsquare~ Location ... (Same as Question A.1)

\noindent \textbf{C.4} Based on the privacy practices I learned purely from the [transparency channel], the [mobile app] (if I install and use it) may share (with other parties) the following types of privacy-sensitive information (Please select all that apply; if no option is applicable, please select “None of above”.):

\noindent\Hsquare~ Location ... (Same as Question A.1)

\noindent \textbf{C.5} Based on the privacy practices I learned purely from the [transparency channel], the [mobile app] (if I install and use it) may use my privacy-sensitive information for the following purposes (Please select all that apply; if no option is applicable, please select “None of above”.):

\noindent\Hsquare~Advertising or marketing
\Hsquare~Fraud prevention and security
\Hsquare~Regulation compliance
\Hsquare~Account management
\Hsquare~App functionality
\Hsquare~Data analytics
\Hsquare~Personalization
\Hsquare~Developer communications
\Hsquare~Others (please specify)
\Hsquare~None of above

\noindent \textbf{C.6} Based on the privacy practices I learned purely from the [transparency channel], the following statements about the data rights that I have regarding my privacy-sensitive information collected by the [mobile app] (if I install and use it) are true (Please select all that apply; if no option is applicable, please select “None of above”.):

\noindent\Hsquare~I can access any privacy-sensitive information collected by the app.
\Hsquare~I can correct any inaccurate privacy-sensitive information held by the app.
\Hsquare~I can have my privacy-sensitive information deleted.
\Hsquare~I can restrict the processing of my privacy-sensitive information.
\Hsquare~I can request a copy of my privacy-sensitive information and have it transferred to another service provider.
\Hsquare~I can object to the processing of my privacy-sensitive information.
\Hsquare~I can request human intervention in the decision-making process that involves my privacy-sensitive information.
\Hsquare~I can lodge my compliant to the supervising authority regarding any misuse of my privacy-sensitive information.
\Hsquare~I can opt-out of the sale or sharing of my privacy-sensitive information.
\Hsquare~Others (please specify)
\Hsquare~None of above

\noindent \textbf{C.7} Based on the privacy practices I learned purely from the [transparency channel], the following statements about the storage or retention of my privacy-sensitive information collected by the [mobile app] (if I install and use it) are true (Please select all that apply; if no option is applicable, please select “None of above”.):

\noindent\Hsquare~The storage location (e.g., on my smartphone, on the app’s remote server, or on a third-party server) is provided to me.
\noindent\Hsquare~The storage or retention period (e.g., six months, two years, or forever) is provided to me.
\Hsquare~The criterion for determining the storage or retention period is provided to me (e.g., my privacy-sensitive information will be stored until my request is addressed by a service representative).
\Hsquare~The post-storage or retention period policy is provided to me (e.g., my privacy-sensitive information will be completely deleted, partially deleted, or anonymized after the retention period).
\Hsquare~Others (please specify)
\Hsquare~None of above

\noindent \textbf{C.8} Based on the privacy practices I learned purely from the [transparency channel], the collection, use, and potentially sharing (with other parties) of my privacy-sensitive information by the [mobile app] (if I install and use it) may harm me by incurring some online risks such as restraining me from expressing my views (e.g., political views) free from worry, allowing attackers to steal my identity, and/or decreasing my chances of getting a good job in the future.

\noindent $\circ$ Strongly agree ... $\circ$ Strongly disagree

\noindent \textbf{C.9} Based on the privacy practices I learned purely from the [transparency channel], the collection, use, and potentially sharing (with other parties) of my privacy-sensitive information by the [mobile app] (if I install and use it) may harm me by incurring some physical risks (e.g., stalking and burglary) to me or my family.

\noindent $\circ$ Strongly agree ... $\circ$ Strongly disagree

\noindent \textbf{C.10} Based on the privacy practices I learned purely from the [transparency channel], the collection, use, and potentially sharing (with other parties) of my privacy-sensitive information by the [mobile app] (if I install and use it) may benefit me by providing me with the customized service tailored to my preferences or needs.

\noindent $\circ$ Strongly agree ... $\circ$ Strongly disagree

\noindent \textbf{C.11} Based on the privacy practices I learned purely from the [transparency channel], the collection, use, and potentially sharing (with other parties) of my privacy-sensitive information by the [mobile app] (if I install and use it) may benefit me by providing me with the enhanced online security protection such as better fraud detection or prevention.

\noindent $\circ$ Strongly agree ... $\circ$ Strongly disagree

\noindent \textbf{C.12} Based on the privacy practices I learned purely from the [transparency channel], I am concerned with the overall privacy risks associated with installing and using the [mobile app].

\noindent $\circ$ Strongly agree ... $\circ$ Strongly disagree

\noindent Please explain your answer to Question C.12 with at least 15 words and with high quality.

\noindent \textbf{C.13} The [transparency channel] should be improved in some ways to help me better understand the privacy practices of the [mobile app] and better estimate the privacy risks associated with installing and using the app.

\noindent $\circ$ Strongly agree ... $\circ$ Strongly disagree

\noindent Please explain your answer to Question C.13 with at least 15 words and with high quality.

\subsection{Post-Interaction Questions}
\label{sec:appendix_Post-Interaction Questions}

\noindent \textbf{D.1} Please rank the three privacy transparency channels in order of the amount of information they provided about the privacy practices of the [mobile app] (1=the channel with the largest amount of information, 3=the channel with the smallest amount of information).

\noindent Data safety: \hspace{14.5mm} $\circ$  1	 $\circ$ 2	$\circ$  3\\
\noindent Privacy policy: \hspace{10.7mm} $\circ$ 1	$\circ$ 2 $\circ$ 3\\
\noindent Permission manifest: \hspace{3.2mm} $\circ$ 1	$\circ$ 2 $\circ$ 3

\noindent Please explain your answer to Question D.1 with at least 15 words and with high quality (in addition, you can also indicate if some of those channels are tied in this ranking).

\noindent \textbf{D.2} Please rank the three privacy transparency channels in order of the clarity of the information they provided about the privacy practices of the [mobile app] (1=the channel with the best information clarity, 3=the channel with worst information clarity).

\noindent Data safety: ... (Same as Question D.1, with explanation)

\noindent \textbf{D.3} Please rank the three privacy transparency channels in order of the intuitiveness of their user interfaces in providing the privacy practices of the [mobile app] (1=the most intuitive channel, 3=the least intuitive channel).

\noindent Data safety: ... (Same as Question D.1, with explanation)

\noindent \textbf{D.4} Please rank the three privacy transparency channels in order of the their overall effectiveness in helping you understand the privacy practices of the [mobile app] and estimate the privacy risks associated with installing and using it (1=the most effective channel, 3=the least effective channel).

\noindent Data safety: ... (Same as Question D.1, with explanation)

\noindent \textbf{D.5} I hope that in the future the privacy transparency channels can be significantly improved in the following ways to help me better understand the privacy practices of mobile apps and better estimate the privacy risks associated with installing and using them (Please select all that apply; if no option is applicable, please select “None of above”.):

\noindent\Hsquare~Provide more and clearer information about an app’s privacy practices.
\Hsquare~Improve their user interfaces to present an app’s privacy practices more effectively.
\Hsquare~Provide a single interface to present their combined information about an app’s privacy practices.
\Hsquare~Highlight the potential privacy risks associated with an app’s privacy practices.
\Hsquare~Make it mandatory for me to interact with them before allowing me to install an app.
\Hsquare~Others (please specify)
\Hsquare~None of above

\noindent\textbf{D.6} Other comments, suggestions, and/or concerns about the privacy transparency channels and/or this study?

\subsection{Post-Study Questions}
\label{sec:appendix_Demographic Questions}

\noindent \textbf{E.1} What is your gender?

 \noindent $\circ$~Female
 $\circ$ Male
 $\circ$ Non-binary
 $\circ$ Prefer to self describe: (Please specify)
 $\circ$ Prefer not to answer

\noindent \textbf{E.2} What is your age group?

 \noindent $\circ$ Under 18 years old
 $\circ$ 18-24 years
 $\circ$ 25-39 years
 $\circ$ 40-49 years
 $\circ$ 50-59 years
 $\circ$ 60 years or above
 $\circ$ Prefer not to answer

\noindent \textbf{E.3} What is your highest education level?

 \noindent$\circ$ No high school degree
 $\circ$ High school degree
 $\circ$ Associate degree
 $\circ$ Bachelor's degree
 $\circ$ Master's degree
 $\circ$ Doctorate degree
 $\circ$ Medical degree
 $\circ$ Others (please specify)
$\circ$ Prefer not to answer

\noindent \textbf{E.4} What is your occupation?

\section{Additional Design Details and Results}
\label{sec:appendix_Results and analysis}
\subsection{Additional Design Details}
\label{sec:appendix_Additional Design Details}

\textit{\textbf{More details about app selection regarding estimating apps’ privacy practice disclosure extensiveness.}}
To be accurate, we manually count the number of relevant entries 
in the ``Data shared'', ``Data collected'', and ``Security practices'' sections of each app's data safety channel; 
we then read the app's privacy policy to count the number of privacy practices
mentioned in it; we further count the number of relevant system permissions~(e.g., the GPS location and camera)
that the app may require to access.
We normalize these three types of numbers, respectively, to values between 0 and 1 for all 20 apps in a category, and add the three normalized values for each app as the estimation of its privacy practice disclosure extensiveness score.  
We further choose the two apps with the lowest and the highest scores in a category as the apps with minimal and extensive disclosures, respectively.

\textbf{\textit{More details about the design of the data safety channel interfaces.}} 
Figure~\ref{fig:interface_data_safety} in Appendix~\ref{sec:appendix_screenshots} exemplifies our simulated interfaces for the data safety channel of the USA TODAY app.
Figure~\ref{fig:interface_data_safety_A} is the interface for the app's data safety summary section on its main page.  The ``See details'' link or the right arrow symbol can take a participant to the interface for the app's detailed data safety page (Figure~\ref{fig:interface_data_safety_B}).Information on the data safety page is organized into sections that correspond to different types privacy and security practices (e.g., ``Data shared'', ``Data collected''). 
Identical to the design of the Google Play Store, our website conceals the purposes of each practice within a collapsible component; it utilizes an instruction prompt to guide participants in expanding these components to access all information.
Furthermore, we add checkboxes alongside the section titles,
asking participants to explicitly indicate that they have read each section. 
The checkboxes require minimal interaction effort from the participants, and they are incorporated into all interfaces of the three channels. 

\textbf{\textit{More details about the design of the privacy policy channel interface.}}
Figure~\ref{fig:interface_privacy_policy} in  Appendix~\ref{sec:appendix_screenshots} exemplifies our simulated interface for the privacy policy channel of the USA TODAY app.
We obtain an app's privacy policy by accessing the URL provided on the Google Play Store.
We present the condensed content of a policy in the privacy policy channel interface as mentioned above, but keep the structure and sentences of each original policy. 

\textbf{\textit{More details about the design of the permission manifest channel interface.}}
Figure~\ref{fig:interface_permission_manifest} in Appendix~\ref{sec:appendix_screenshots} exemplifies our simulated interface for the permission manifest channel of the USA TODAY app.
Permissions are categorized (e.g., location, contacts) based on their capabilities or functionalities, and are briefly described 
(e.g., a description [access approximate location (network-based)] is given under the ``Location'' permission). 

\textbf{\textit{Pilot study.}}
We recruited 20 participants from Prolific in a pilot study to evaluate the smoothness of our study procedure, the clarity of our questions, and the implementation of our website. 
We observed no major issues, and only made minor adjustments to the formal study by 
removing a redundant question, 
relaxing the word count requirement for open-ended responses from 20 to 15 as extra words provided little information,
and adjusting the compensation structure to include a \$2 USD bonus for participants with high-quality responses. 
The pilot study participants and their responses were excluded from our formal study and result analysis.
Note that the average and median time for completing the pilot study was 48 and 39 minutes, respectively.

\subsection{Additional Results}
\label{sec:appendix_Additional Results}
\begin{figure}[h]
  \centering
  \includegraphics[width=0.5\textwidth]{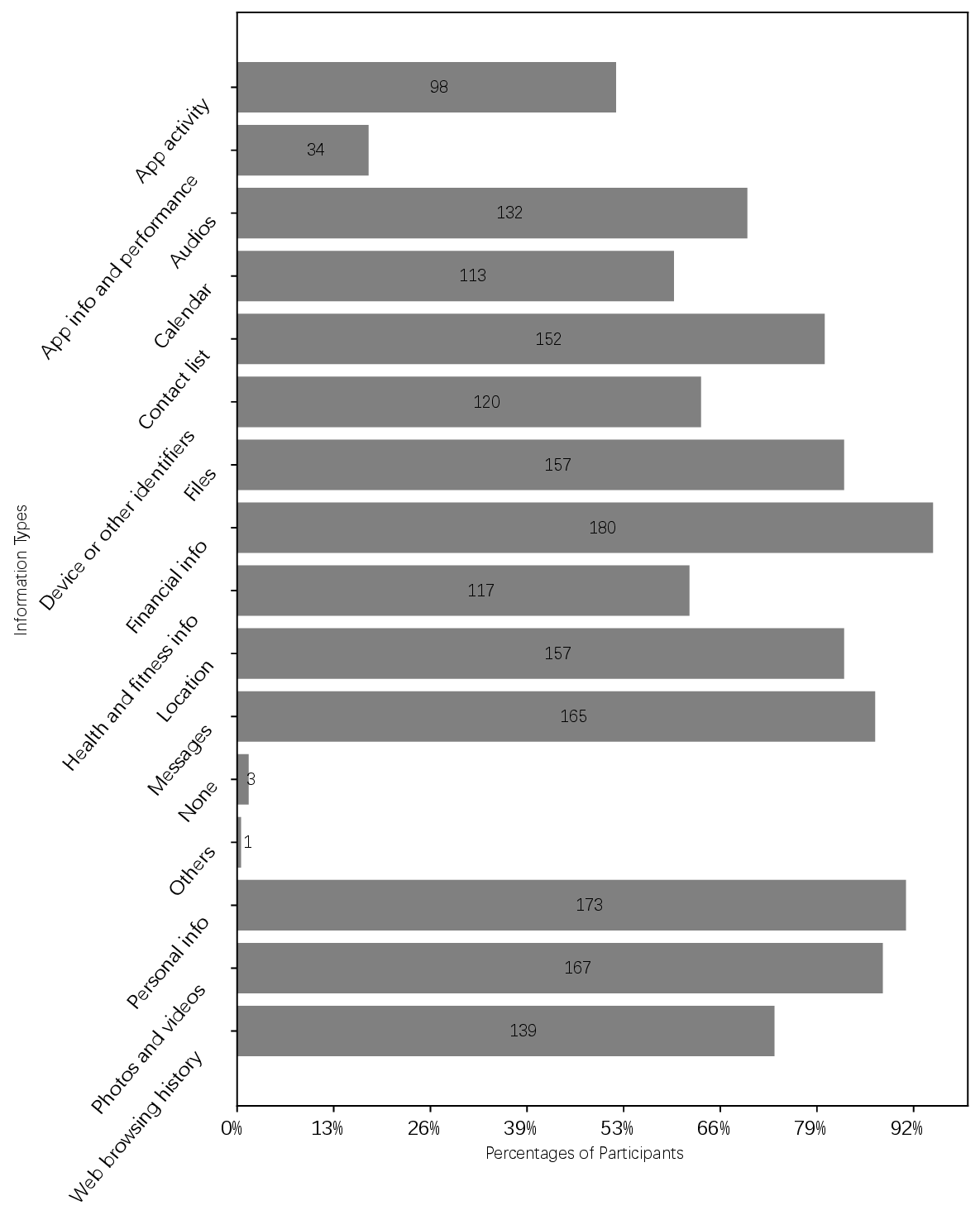}
   \vspace{-15pt}
  \caption{Percentages (and Numbers) of Participants Who Considered Some Types of Information as Privacy-Sensitive (Question A.1, n=190)}
  \label{fig:A1}
\end{figure}

\begin{figure}[h]
    \centering
    \includegraphics[scale=0.5]{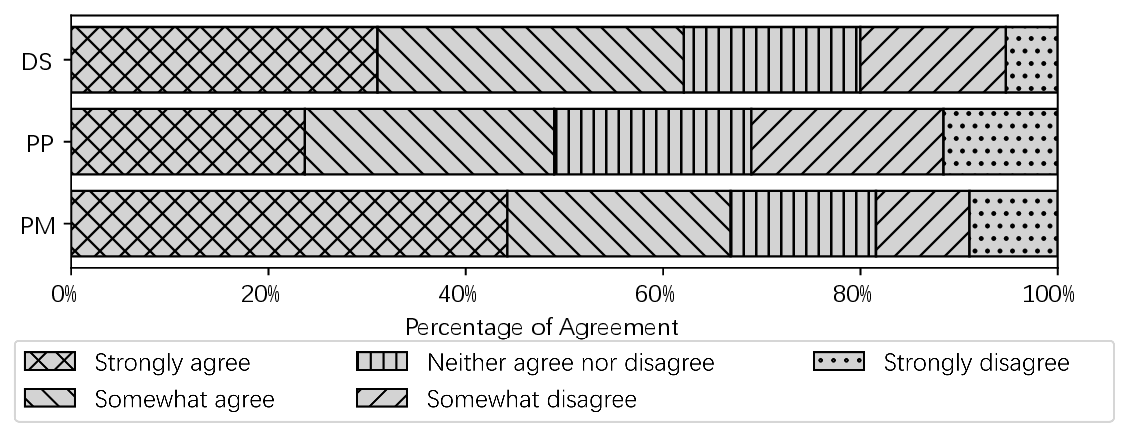}
\caption{Participants' Agreement Levels on If a Particular Channel Should be Improved (Questions C.13, n=190).}
\label{fig:C13}
\end{figure}

Table~\ref{tab:demographic} shows participants' demographic distributions.
Figure~\ref{fig:A1} illustrates the percentages (and numbers) of participants who considered some types of information as privacy-sensitive. 
Table~\ref{tab:RQ2_participants_perceptions} shows the 190 participants’ perceived risks, benefits, and concerns.
Figure~\ref{fig:C13} shows participants' agreement levels on if a particular channel should be improved.

\begin{table}[h!]
\centering
\small
  \caption{Demographics of Participants (n=190)}
  \label{tab:demographic}
  \begin{tabular}{ll}
    \toprule
    & Numbers (and \%) of Participants  \\
    \midrule
    \textbf{Gender (Question E.1)} &  \\
    Male & 93~(48.9\%) \\
    Female & 90~(47.4\%)\\
    Non-binary & 7~(3.7\%) \\
    \hline
    \textbf{Age (Question E.2)}\\
    18-24 & 12~(6.3\%) \\
    25-39 & 88~(46.3\%) \\
    40-49 & 46~(24.2\%) \\
    50-59 & 28~(14.7\%) \\
    60-more & 15~(7.9\%) \\
    Preferred Not To Answer & 1~(0.5\%)\\
    
    \hline
    \textbf{Education (Question E.3)} & \\
    Bachelor's Degree & 70~(36.8\%) \\
    High School Degree & 66~(34.7\%) \\
    Associate Degree & 29~(15.3\%) \\
    Master's Degree & 16~(8.4\%) \\
    Doctorate Degree & 3~(1.6\%) \\
    Preferred Not To Answer & 3~(1.6\%) \\
    Medical Degree & 1~(0.5\%) \\
    No High School Degree & 1~(0.5\%) \\
    Vocational Education & 1~(0.5\%) \\
    
    \hline
    \textbf{Occupation (Question E.4)}  & \\
    Unemployed and Retired & 40~(21.1\%) \\
    Information Technology & 23~(12.1\%) \\
    Management and Administration & 21~(11.1\%) \\
    Freelance and Self-Employed & 16~(8.4\%) \\
    Retail and Sales & 16~(8.4\%) \\
    Other Specific Professions & 15~(7.9\%) \\
    Creative and Design & 11~(5.8\%) \\
    Finance & 8~(4.2\%) \\
    Healthcare and Medical & 7~(3.7\%) \\
    Manufacturing and Production & 7~(3.7\%) \\
    Transportation & 6~(3.2\%) \\
    Teacher & 6~(3.2\%) \\
    Student & 4~(2.1\%) \\
    Prefer not to answer & 3~(1.6\%) \\
    Admin Support & 2~(1.1\%) \\
    Real Estate & 2~(1.1\%) \\
    Construction & 1~(0.5\%) \\
    Paralegal & 1~(0.5\%) \\
    Athletes & 1~(0.5\%) \\
  \bottomrule
\end{tabular}

\end{table}

\begin{table*}
\centering
\small
  \caption{
The 190 Participants' Perceived Risks, Benefits, and Concerns. 
Participant percentage changes are calculated between the responses before (in Questions B.3 to B.7) and after (in Questions C.8 to C.12) the interaction with each privacy transparency channel: DS (Data Safety), PP (Privacy Policy), and PM (Permission Manifest).
}.
  \label{tab:RQ2_participants_perceptions}
\scalebox{0.82}{
  \begin{tabular}{lllllll}
    \toprule
     \textbf{Brief Content of Questions} & \textbf{Question\#} & \makecell[l]{\textbf{Strongly} \\ \textbf{Agree}}  &  \textbf{Agree} &  \textbf{Neutral} & \textbf{Disagree}  &  \makecell[l]{\textbf{Strongly} \\ \textbf{Disagree}} 
    \\
    \midrule
    \makecell[l] {
    Based on what I learned from the\\ 
    \{app description\} or \{privacy transparency channel\},\\ 
    the privacy practices of the assigned app\\
    may harm me by incurring some \textbf{online risks}.
    }
    & 
    \makecell[l] {
     Baseline (B.3)\\
    DS (C.8)\\
    PP (C.8)\\
    PM (C.8)
     } 
     &
    \makecell[l] {
     6.3\%\\
    11.1\% (+4.8\%)\\
    11.6\% (+5.3\%)\\
    12.6\% (+6.3\%)
    }
    &
    \makecell[l] {
     17.4\%\\
    18.9\% (+1.5\%)\\
    27.9\% (+10.5\%)\\
    24.2\% (+6.8\%)
    }
    &
    \makecell[l] {
     26.8\%\\
    20.5\% (-6.3\%)\\
    14.7\% (-12.1\%)\\
    18.9\% (-7.9\%)
    }
    &
    \makecell[l] {
     26.8\%\\
    16.3\% (-10.5\%)\\
    17.9\% (-8.9\%)\\
    18.4\% (-8.4\%)
    }
    &
    \makecell[l] {
     22.6\%\\
    33.2\% (+10.6\%)\\
    27.9\% (+5.3\%)\\
    25.8\% (+3.2\%)
    }\\
    
    \hline

    \makecell[l] {
    Based on what I learned from the\\ 
    \{app description\} or \{privacy transparency channel\},\\ 
    the privacy practices of the assigned app\\
    may harm me by incurring some \textbf{physical risks}.
    } 
    & 
    \makecell[l] {
     Baseline (B.4)\\
    DS (C.9)\\
    PP (C.9)\\
    PM (C.9)
     } 
     &
    \makecell[l] {
     6.3\%\\
    10.0\% (+3.7\%)\\
    10.0\% (+3.7\%)\\
    8.9\% (+2.6\%)
    }
    &
    \makecell[l] {
     14.2\%\\
    13.7\% (-0.5\%)\\
    14.2\% (0.0\%)\\
    19.5\% (+5.3\%)
    }
    &
    \makecell[l] {
     20.5\%\\
    15.8\% (-4.7\%)\\
    17.4\% (-3.1\%)\\
    21.6\% (+1.1\%)
    }
    &
    \makecell[l] {
     26.3\%\\
    16.8\% (-9.5\%)\\
    23.7\% (-2.6\%)\\
    19.5\% (-6.8\%)
    }
    &
    \makecell[l] {
     32.6\%\\
    43.7\% (+11.1\%)\\
    34.7\% (+2.1\%)\\
    30.5\% (-2.1\%)
    }\\
    
    \hline
    
    \makecell[l] {
    Based on what I learned from the\\ 
    \{app description\} or \{privacy transparency channel\},\\ 
    the privacy practices of the assigned app\\
    may benefit me with the \textbf{customized service}.
    }
    & 
    \makecell[l] {
     Baseline (B.5)\\
    DS (C.10)\\
    PP (C.10)\\
    PM (C.10)
     } 
     &
    \makecell[l] {
     22.1\%\\
    17.9\% (-4.2\%)\\
    20.5\% (-1.6\%)\\
    15.8\% (-6.3\%)
    }
    &
    \makecell[l] {
     47.4\%\\
    37.9\% (-9.5\%)\\
    47.4\% (0.0\%)\\
    31.1\% (-16.3\%)
    }
    &
    \makecell[l] {
     16.3\%\\
    25.3\% (+9.0\%)\\
    18.4\% (+2.1\%)\\
    29.5\% (+13.2\%)
    }
    &
    \makecell[l] {
     8.9\%\\
    8.9\% (0.0\%)\\
    6.3\% (-2.6\%)\\
    11.1\% (+2.2\%)
    }
    &
    \makecell[l] {
     5.3\%\\
    10.0\% (+4.7\%)\\
    7.4\% (+2.1\%)\\
    12.6\% (+7.3\%)
    }\\
    
    \hline
    \makecell[l] {
    Based on what I learned from the\\ 
    \{app description\} or \{privacy transparency channel\},\\ 
    the privacy practices of the assigned app\\
    may benefit me with the \textbf{security protection}.
    }
    & 
    \makecell[l] {
     Baseline (B.6)\\
    DS (C.11)\\
    PP (C.11)\\
    PM (C.11)
     } 
     &
    \makecell[l] {
     4.2\%\\
    10.5\% (+6.3\%)\\
    14.2\% (+10.0\%)\\
    7.4\% (+3.2\%)
    }
    &
    \makecell[l] {
     16.8\%\\
    26.8\% (+10.0\%)\\
    36.8\% (+20.0\%)\\
    20.0\% (+3.2\%)
    }
    &
    \makecell[l] {
     31.6\%\\
    27.9\% (-3.7\%)\\
    26.8\% (-4.8\%)\\
    32.1\% (+0.5\%)
    }
    &
    \makecell[l] {
     24.2\%\\
    15.3\% (-8.9\%)\\
    7.4\% (-16.8\%)\\
    16.8\% (-7.4\%)
    }
    &
    \makecell[l] {
     23.2\%\\
    19.5\% (-3.7\%)\\
    14.7\% (-8.5\%)\\
    23.7\% (+0.5\%)
    }\\
    
    \hline

    \makecell[l] {
    Based on what I learned from the\\ 
    \{app description\} or \{privacy transparency channel\},\\ 
    \textbf{I am concerned} with the overall privacy risks \\
    associated with installing and using the app.
    }
    & 
    \makecell[l] {
     Baseline (B.7)\\
    DS (C.12)\\
    PP (C.12)\\
    PM (C.12)
     } 
     &
    \makecell[l] {
     12.6\%\\
    22.6\% (+10.0\%)\\
    25.3\% (+12.7\%)\\
    32.6\% (+20.0\%)
    }
    &
    \makecell[l] {
     24.7\%\\
    26.8\% (+2.1\%)\\
    21.6\% (-3.1\%)\\
    30.5\% (+5.8\%)
    }
    &
    \makecell[l] {
     18.4\%\\
    11.1\% (-7.3\%)\\
    13.7\% (-4.7\%)\\
    8.9\% (-9.5\%)
    }
    &
    \makecell[l] {
     35.3\%\\
    26.8\% (-8.5\%)\\
    27.4\% (-7.9\%)\\
    15.8\% (-19.5\%)
    }
    &
    \makecell[l] {
     8.9\%\\
    12.6\% (+3.7\%)\\
    12.1\% (+3.2\%)\\
    12.1\% (+3.2\%)
    }\\
    
    \hline
    
  \bottomrule
\end{tabular}
}
\end{table*}

\textbf{\textit{Summary of the Analysis of Response Accuracy for Questions C.3 to C.7 from the Three Aspects.}}
For Questions C.3 to C.7,
the detailed results from the three aspects are provided in Section~2 of the Supplementary Material~\cite{github_link}.
In summary,
the three aspects have a varying impact on participants' response accuracy. 
For example, 
from the app category aspect, only 19.1\% of dating app participants achieved 0.8 or greater accuracy in identifying data-sharing practices
 (Question C.4) from the DS channel, while the corresponding percentage of news app participants is 73.1\%.
From the app disclosure extensiveness aspect, 
53.2\% and 28.5\% of extensive disclosure app participants achieved 0.8 or greater accuracy in identifying data collection and use practices (Question C.3) from the PP and PM channels, respectively, compared to 16.0\% and 52.0\% of minimal disclosure app participants.
From the channel sequence aspect, we found no noticeable differences in response accuracy between the two participant groups for all channels.

\textbf{\textit{Summary of the Analysis of Perceived Risks (Questions B.3 and B.4; C.8 and C.9) and Benefits (Questions B.5 and B.6; C.10 and C.11) from the Three Aspects.}}
The detailed results from the three aspects are provided in Section~3 of the Supplementary Material~\cite{github_link}. 
In summary,
from the app category aspect, we found that the percentages of dating app participants who agreed with the statements on perceived risks, both before and after interacting with each channel, are significantly higher than that of news app participants. 
Meanwhile, dating app participants were more likely to perceive some benefits than news app participants. 
From the app disclosure extensiveness aspect, there is no statistically significant difference between the two participant groups regarding perceived risks based on the app description; there are significantly higher percentages of extensive disclosure app participants who 
agreed with the statements 
compared to minimal disclosure app participants for all channels.
However, we found no statistically significant differences from the channel sequence aspect.

\end{appendices}



\end{document}